\documentclass[aps,prx,amsmath,amssymb,twocolumn,floatfix,longbibliography, 10pt]{revtex4-2}

\usepackage{silence}
\usepackage{graphicx}
\usepackage{dcolumn}
\usepackage{bm}
\usepackage[utf8]{inputenc}
\usepackage[english]{babel}
\usepackage{amsfonts}
\usepackage{booktabs}
\usepackage{hyperref}
\usepackage{physics}
\usepackage{epstopdf}
\usepackage{mathtools}
\usepackage[capitalise]{cleveref}
\usepackage{float}
\usepackage[space]{grffile}
\usepackage{color}
\usepackage[normalem]{ulem}
\usepackage{subfigure}
\usepackage{natbib}
\usepackage{amsmath}

\usepackage{textcomp}
\usepackage{gensymb}
\let\strong\textbf

\newcommand{\appref}[1]{Appendix \ref{#1}}

\makeatletter
\let\@orig@make@capt@title\@make@capt@title
\def\@make@capt@title#1#2{\@orig@make@capt@title{{\bf #1}}{#2}}
\makeatother

\newcommand{\avgz}[1]{\langle #1 \rangle_{z_*(t)}}

\def\diamondcomma{\ \raise.3ex\hbox{$\diamond$}\kern-0.4em\lower.7ex\hbox{$,$}\ }
\def\lesssim{\ \raise.3ex\hbox{$<$}\kern-0.8em\lower.7ex\hbox{$\sim$}\ }
\def\gesim{\ \raise.3ex\hbox{$>$}\kern-0.8em\lower.7ex\hbox{$\sim$}\ }
\DeclareMathOperator{\Real}{Re}

\begin{document}
\title{
Generalized non-reciprocal phase transitions in multipopulation systems
}

\author{Cheyne Weis}
    \affiliation{James Franck Institute and Department of Physics, University of Chicago, Illinois, 60637, USA}
    \email{cheyneweis@gmail.com} 

    \author{Ryo Hanai}
    \affiliation{
    Department of Physics, Institute of Science Tokyo, 2-12-1 Ookayama Meguro-ku, Tokyo, 152-8551, Japan
    }
    \email{hanai.r.7e4b@m.isct.ac.jp} 

\begin{abstract}
        Non-reciprocal interactions are prevalent in various complex systems leading to phenomena that cannot be described by traditional equilibrium statistical physics. 
        Although non-reciprocally interacting systems composed of two populations have been closely studied, the physics of non-reciprocal systems with a general number of populations is not well explored despite the relevance to biological systems, active matter, and driven-dissipative quantum materials.
        In this work, we investigate the generic features of the phases and phase transitions that emerge in
        $O(2)$ symmetric many-body systems with multiple non-reciprocally coupled populations, applicable to microscopic models such as networks of oscillators, flocking models, and more generally systems where each agent has a phase variable. 
        Using symmetry and topology of the possible orbits, we systematically show that a rich variety of time-dependent phases and phase transitions arise.
        Examples include multipopulation chiral phases that are distinct from their two-population counterparts and emerge via a phase transition characterized by critical exceptional points, as well as limit cycle saddle-node bifurcation and Hopf bifurcation. Interestingly, we find a phase transition that dynamically restores the $\mathbb{Z}_2$ symmetry occurs via a homoclinic orbit bifurcation, where the two $\mathbb{Z}_2$ broken orbits merge at the phase transition point, providing a general route to homoclinic chaos in the order parameter dynamics for $N\geq4$ populations. Our framework provides general principles for understanding non-equilibrium heterogeneous systems and guides experimental exploration into such systems.
    \end{abstract}

\date{\today}

\maketitle

\section{Introduction}
Processes fundamental to everyday experience — whether they drive biological activity, enable cognition, or maintain ecosystems — are inherently out of equilibrium to sustain the dynamics essential for life, computation, and complex adaptive behaviors.
A key ingredient for many types of non-equilibrium dynamics is non-reciprocity, where for two agents A and B, the interaction from A to B is different from the interaction from B to A. Networks of neurons~\cite{wilsonExcitatoryInhibitoryInteractions1972}, ecosystems~\cite{rosenzweigGraphicalRepresentationStability1963}, chemical reaction networks~\cite{lotkaAnalyticalNoteCertain1920}, and social interactions~\cite{nowakEvolutionaryGamesSpatial1992} often exhibit non-reciprocity through predator-prey dynamics, where one population's interests are diametrically opposed to the other's, creating oscillatory dynamics or complex cascades of activity. 
Recent explorations have found collective phenomena and exotic non-equilibrium phases of matter that accompany non-reciprocity, such as long-range order in two spatial dimensions~\cite{Loos2023, Dadhichi2020, Pisegna2024}, odd viscosity and elasticity~\cite{Fruchart2023,soniOddFreeSurface2019,markovichNonreciprocityOddViscosity2024, scheibnerOddElasticity2020},  
time-crystalline order~\cite{huangActivePatternFormation2024, hanaiNonreciprocalFrustrationTime2024, Fruchart2021, youNonreciprocityGenericRoute2020, Saha2020}, 
non-equilibrium boundary modes~\cite{muruganTopologicallyProtectedModes2017}, effects of spatial inhomogeneity~\cite{weiderpassSolvingKineticIsing2025, godrecheDynamicsDirectedIsing2011, searaNonreciprocalInteractionsSpatially2023, veenstraNonreciprocalTopologicalSolitons2024, belyanskyPhaseTransitionsNonreciprocal2025, dasDrivenHeisenbergMagnets2002, bhattEmergentHydrodynamicsNonreciprocal2023}, and chaotic phases~\cite{parkavousiEnhancedStabilityChaotic2024}, 
with applications in active matter~\cite{bowickSymmetryThermodynamicsTopology2022}, biology~\cite{zhengTopologicalMechanismRobust2024, tanOddDynamicsLiving2022}, neural networks~\cite{sompolinskyTemporalAssociationAsymmetric1986}, and robotic metamaterials~\cite{brandenbourgerNonreciprocalRoboticMetamaterials2019}. 
Analogous non-reciprocal collective phenomena are also of great interest in 
open quantum systems~\cite{khassehActiveQuantumFlocks2024,nadolnyNonreciprocalSynchronizationActive2024,chiacchioNonreciprocalDickeModel2023, hanaiPhotoinducedNonreciprocalMagnetism2024}.

A generic feature introduced by non-reciprocal interactions is the emergence of persistent time-dependent many-body phases~\cite{hanaiNonreciprocalFrustrationTime2024, Fruchart2021, youNonreciprocityGenericRoute2020, Saha2020, zelleUniversalPhenomenologyCritical2024} prohibited at equilibrium~\cite{watanabeAbsenceQuantumTime2015}.
Examples include traveling waves~\cite{braunsNonreciprocalPatternFormation2024} that can occur in lipid domains~\cite{johnTravellingLipidDomains2005} and active-passive Brownian mixtures~\cite{Saha2020}, as well as non-equilibrium systems in biology~\cite{demirDynamicsPatternFormation2020}, quantum optics~\cite{chiacchioNonreciprocalDickeModel2023,nadolnyNonreciprocalSynchronizationActive2024}, and driven condensates~\cite{Hanai2019}.
Rather uniquely, phase transitions induced by non-reciprocity can occur at critical exceptional points, where the mode that exhibits critical slowing aligns with an existing zero mode from a broken $SO(2)$ symmetry~\cite{Fruchart2021,Hanai2020}.

Most works in the literature have focused on non-reciprocally coupled one~\cite{huangActivePatternFormation2024}
or two-population systems~\cite{Fruchart2021,avniNonreciprocalIsingModel2024b,Saha2020}.
However, systems in nature often contain diverse subpopulations. Examples of such systems include neural networks with multiple types of inhibitory neurons~\cite{faugerasConstructiveMeanfieldAnalysis2009,kimDynamicsMultipleInteracting2020, palmigianoCommonRulesUnderlying2023}, ecological networks~\cite{reichenbachMobilityPromotesJeopardizes2007, blumenthalPhaseTransitionChaos2024,huEmergentPhasesEcological2022, allesinaStabilityCriteriaComplex2012}, socioeconomic models with the addition of a population of a ``committed minority"~\cite{xieSocialConsensusInfluence2011}, and multicomponent conserved densities~\cite{parkavousiEnhancedStabilityChaotic2024}, often arising in biological settings~\cite{jacobsPhaseTransitionsBiological2017, jacobsSelfAssemblyBiomolecularCondensates2021}. 
Recent studies in such biological settings have found that although there can be thousands of active components, the system can often be described by a small number of effective components~\cite{grafThermodynamicStabilityCritical2022, chenEmergenceMultiphaseCondensates2024}.
This motivates the study of systems with a moderate number of populations. 

In this paper, we investigate the phases and phase transitions that emerge in non-reciprocally interacting systems composed of three or more populations ($N\ge 3)$.
For $N=2$ population non-reciprocal systems, the time-dependent trajectories are generically limit cycles — periodic orbits with discrete time translational symmetry, sometimes referred to as a time-crystalline phase~\cite{khemaniBriefHistoryTime2019, kongkhambutObservationContinuousTime2022}. However, multipopulation systems are expected to exhibit a greater number of topologically distinct limit cycles or alternative phases with more complex dynamics because they exist in a higher dimensional space with a wider variety of possible bifurcations.

\begin{figure*}
    \centering
    \includegraphics[width=\textwidth]{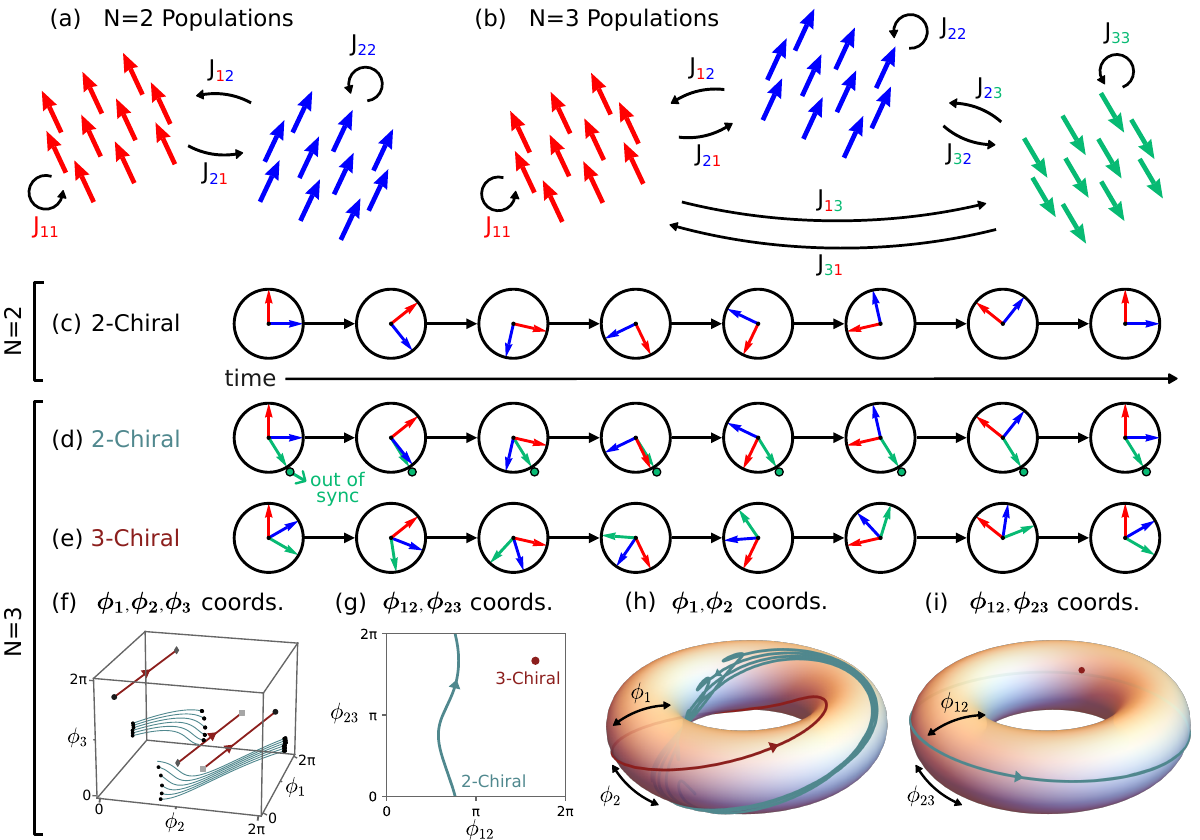}
    \caption{\strong{Topology of the 2 and 3-chiral phases for multiple populations.}
    (a-b) A multipopulation non-reciprocal system for two and three populations. Members within each population, indicated by color, interact with a ferromagnetic reciprocal coupling ($J_{aa}>0$) causing them to align. The coupling between members of different populations is in general \emph{non-reciprocal} ($J_{ab}\neq J_{ba}$).
    (c) The two population chiral phase, where the direction population 1 chases population 2 is spontaneously broken as studied in Ref.\cite{Fruchart2021}.
    (d-e) Schematic snapshots of the 2-chiral and 3-chiral phases for three populations respectively. When the third population is completely decoupled, as seen in panel (d), two populations go around once per period  while one population remains unsynchronized, whereas in the 3-chiral phase (e), all three populations go around once per period. Each chiral phase is topologically distinct, implying that transitioning between different chiral phases requires the system to undergo a phase transition. 
    (f) The behavior of phases $\phi_a$ where $R_ae^{i\phi_a}=\frac{1}{N_a}\sum_i e^{i\theta^a_i}$ is the average phase of members $\theta_i$ of the population in the 2-chiral (maroon)  and 3-chiral phase (teal). Here, we set $J_{12}= 1.77,\ J_{13}=-0.52,\ J_{21}= -1.81,\ J_{23}=0.59,\ J_{31}=0.45,\ J_{32}=-0.51$ for the maroon line and $J_{12}=J_{13} = J_{23}=-J_{21}=-J_{31} = -J_{32}=1$ for the teal line, with  $J_{aa}=10$ and $\Delta=0.75$.
    (g) The 2-chiral orbit is periodic in $\phi_{ab}\coloneqq\phi_a-\phi_b$ coordinates.
    (h-i) compares projections on $\mathbb{T}^2$ of the 2-chiral and 3-chiral state trajectories with left torus giving the $(\phi_1,\phi_2)$ projection and the right torus gives the $(\phi_{12},\phi_{23})$ projection.
    }
    \label{fig:gen_chiral}
\end{figure*}

We illustrate this by comparing the possible phases for two and three populations in $O(2)$ symmetric models (see Fig. \ref{fig:gen_chiral}(a),(b)).
In the two-population case, the system exhibits a non-reciprocal phase transition to the ``chiral phase" \cite{Fruchart2021} (which we refer to as the 2-chiral phase in this paper), where population A chases population B or vice versa, depending on the direction of the non-reciprocal coupling (see Fig.~\ref{fig:gen_chiral}(c)). 
In the three-population case, there exist time-dependent phases distinct from the 2-chiral phase. In addition to population A chasing population B while population C is unsynchronized (see Fig.~\ref{fig:gen_chiral}(d)) \footnote{By unsynchronized, we mean that two populations are oscillating at incommensurate frequencies with one another while the members within each population are synchronized.}, there is a completely synchronized dynamical phase where population A chases population B and population B chases population C (which we refer to as the 3-chiral phase (see Fig.~\ref{fig:gen_chiral}(e)). The novel 3-chiral phase can only occur for $N\geq3$ populations. Importantly, these dynamical phases are topologically distinct (see Fig. \ref{fig:gen_chiral} (f)-(i)), due to the periodic orbits winding in different manners around the torus meaning the system must pass through a phase transition to go between the two behaviors. 
This observation strongly suggests the emergence of richer phases and phase transitions in multipopulations systems beyond the two-component non-reciprocal phase transitions~\cite{Fruchart2021}. 

By systematically exploring the multipopulation dynamic phases and phase transitions in $O(2)$ symmetric systems by examining the constraints on the possible dynamics from the symmetry and topology of the system, we find a rich variety of bifurcations takes place between different dynamical phases.
This includes bifurcations that are accompanied by generalized critical exceptional points~\cite{weisExceptionalPointsNonlinear2023a} where the covariant Lyapunov (Floquet) vectors coalesce at the transition point, as well as global bifurcations such as a symmetric homoclinic orbit bifurcation. 
A novel form of symmetry breaking takes place at the corresponding phase transition. While the $\mathbb{Z}_2$ symmetry is broken by the instantaneous dynamics, the symmetry is `dynamically restored'; 
applying the $\mathbb{Z}_2$ operation and time evolving by one-half of the limit cycle period leaves the trajectory invariant. 
This is in stark contrast to the conventional spontaneous symmetry breaking (such as those occurring in the ferromagnetic phase of an Ising model), where the $\mathbb{Z}_2$ operation swaps the two $\mathbb{Z}_2$-symmetry broken attractors, as seen in panels (b-c) in Fig. \ref{fig:bifurcations}.
We find that periodic orbits can be systematically understood from symmetry-based arguments combined with bifurcation theory. 
Our analysis is then extended to systems with larger numbers of populations, allowing for additional homoclinic orbit limit cycle bifurcations, quasiperiodic attractors, and chaotic phases induced at strong non-reciprocity,  which further break the discrete time translational symmetry that exists for the limit cycle. 
We discuss the topological classification of quasiperiodic attractors, and the emergence of chaos near the $\mathbb{Z}_2$-symmetry restoring phase transition. 

The paper is organized as follows. In Sec.~\ref{sec: model}, we introduce the $N-$population model of non-reciprocal agents. In Sec.~\ref{sec:topology}, we give the framework for the classification of periodic phases in general models from topology and symmetry, and demonstrate its efficacy in the three population scenario. In Sec.~\ref{sec:higherpop}, we explore additional phenomena that arise for arbitrary numbers of populations. 

\section{Model} 
\label{sec: model}

We consider the generic order parameter $z_a = R_ae^{i\phi_a}\in \mathbb{C}$ dynamics of $O(2)$ models with multiple populations indexed by $a$ each having many individual agents.
The most general form of the order parameter dynamics for such a system is given by,
\begin{equation}
    \dot{z}_a = 
    \sum_b J_{ab}z_b - \sum_{b,c,d}
    K_{abcd}\bar{z}_b z_c z_d
    +O(z^5),
    \label{eq:eft}
\end{equation} within a mean-field approximation and with real parameters $J_{ab}$ and $K_{abcd}$.
Although our arguments based on symmetry and topology will apply to the general case Eq.~\eqref{eq:eft}, as a concrete example, we consider the Ott-Antonsen equation \cite{Ott2002}
\begin{equation}
    \dot{z}_a = -\Delta z_a+\frac{1}{2}\sum_b
    J_{ab}(z_b - z_a^2 \bar{z}_b).
    \label{eq:OA}
\end{equation}
Eq. \eqref{eq:OA} exactly describes the order parameter dynamics of the Kuramoto model \cite{Kuramoto1984,Acebron2005} for coupled oscillators,
\begin{equation}
    \label{eq:kuramoto}
        \dot{\theta}_i^a = \omega_i^a + \sum_{b}\sum_{j=1}^{N_b}\frac{J^{ij}_{ab}}{N_b}\sin(\theta^b_j-\theta^a_i)
\end{equation}
used as a model of neural\cite{ashwinMathematicalFrameworksOscillatory2016,bickUnderstandingDynamicsBiological2020,weerasinghePredictingEffectsDeep2019} and ecological systems\cite{vandermeerNewFormsStructure, gironSynchronizationUnveilsOrganization2016a} as well as Josephson arrays\cite{wiesenfeldFrequencyLockingJosephson1998}. The order parameter, defined as $z_a=\frac{1}{N_a}\sum_i e^{i\theta^a_i}$, characterizes the strength of the synchronization.
The quenched frequency disorder $\omega_i^a$ is assumed to be drawn from a Cauchy-Lorentz distribution with width $\Delta$, i.e., $p_i^a(\omega_i^a)=(\Delta/\pi) /[(\omega_i^a)^2+\Delta^2]$.
In fact, as shown in \appref{app:models}, the order parameter dynamics of a number of models reduce to Eq. \eqref{eq:OA} in the limit of small ``disorder" $\Delta$. Examples of such models include non-reciprocal XY models, disordered oscillator models, and topological flocking models,
where disorder $\Delta$ represents noise strength for non-reciprocal XY spins, the frequency distribution variance for arbitrarily disordered oscillators, and the interaction range for self-propelled particles, respectively. 


We restrict ourselves in this paper to the case where each population is in the ordered phase (i.e. $|z_a|>0$). In this regime, the slow dynamics primarily occur in the phase $\phi_a$ of each population, where the amplitude $|z_a|$ adiabatically follows the dynamics of the phase, i.e. $\partial_t |z_a|\approx 0$. 

\section{Topology and symmetry of dynamic phases}
\label{sec:topology}
We now examine the possible phases and phase transitions that can occur in the multipopulation $O(2)$-symmetric systems introduced above. 
Symmetry --- and its implications for the possible symmetry-breaking patterns in the $O(2)$ model --- give strong constraints to help classify phases.
Recall that the action of reflections on $SO(2)$ inverts the angle of a rotation, yielding the semidirect product structure $O(2) \cong SO(2) \rtimes \mathbb{Z}_2$,
where the action of $SO(2)$ here is $z_a\rightarrow e^{i\alpha}z_a$ for arbitrary $\alpha\in\mathbb{R}$, while the action of $\mathbb{Z}_2$ is $z_a\rightarrow \overline{z}_a$. The transition from the disordered ($z_a=0$) to the ordered ($z_a\ne 0$) phase is associated with the spontaneous $SO(2)$ symmetry breaking, leaving the system symmetric under the $\mathbb{Z}_2$ symmetry in the ordered phase. 
As already mentioned, non-reciprocity can result in chase-and-runaway dynamics between different populations, often causing an additional spontaneous breaking of the $\mathbb{Z}_2$ reflection symmetry. 
Importantly, the underlying $\mathbb{Z}_2$ symmetry of the equations of motion Eq. \eqref{eq:eft} implies that for every solution $z_a(t)$ of Eq.~\eqref{eq:eft}, the $\mathbb{Z}_2$ conjugate state $\bar z_{a*}(t)$ must also be a solution of Eq.~\eqref{eq:eft}. 
For example, the 2-chiral phase spontaneously breaks the $\mathbb{Z}_2$ symmetry exhibiting a counterclockwise $z_a^{\rm CCW}(t)\propto e^{i\Omega t}$ or clockwise rotation $z_a^{\rm CW}(t)\propto e^{-i\Omega t}$. (See Fig.~\ref{fig:gen_chiral}(c).) The two families of solutions are related to each other by the $\mathbb{Z}_2$-symmetry operation $z_a^{\rm CW}(t)\sim\bar z_a^{\rm CCW}(t)$, where we use $\sim$ to mean the solutions are equivalent up to a rotation.  

Given the above constraints and symmetry-breaking patterns, we now examine the possible topology of the orbits by first considering the case of stable limit cycle solutions (i.e. $z_a(t) = z_a(t+T)$ for a period $T$). The phase space manifold of the variables $\phi_a$ in Eq. \eqref{eq:OA} is equivalent to the $N$-torus ($\mathbb{T}^N$). The compact space $\mathbb{T}^N$ arises naturally as the phase space of ordered non-reciprocal matter studied here with a limit cycle simply being a closed orbit in this compact space.
Any closed orbit on $\mathbb{T}^N$ can be assigned an $N$-tuple of integers that represent how many times the curve wraps around each dimension of the torus and is unchanged by any isotopy (smooth deformation of the curve connected to the identity) \cite{hatcherAlgebraicTopology2001, frankelGeometryPhysicsIntroduction2012, bredonTopologyGeometry2005}.

To characterize the various orbits, it is useful to assign the winding number $w_a$ via the integral
\begin{equation}
    \label{eq:winding_a}
    w_a = \frac{1}{2\pi}\int_{0}^{T} \dot\phi_{a}(t)dt,
\end{equation}
which can be safely defined for periodic trajectories with period $T$. An example of a system with a non-trivial winding $w_a\ne 0$ is the 2-chiral phase appearing in a two-population system.
As mentioned earlier, the 2-chiral phase has orbits with counter-clockwise (clockwise) rotation $z_a^{\rm CCW}=|z_a|e^{i\Omega t}$ ($z_a^{\rm CW}=|z_a|e^{-i\Omega t}$),  implying $\phi^{CCW}_a(t) = \Omega t\ \ (\phi^{CW}_a(t) =-\Omega t)$.
This trajectory has the winding numbers $(w_1, w_2) = (1,1)$ ($(w_1, w_2)=(-1,-1)$), meaning both populations wind once per orbit in both of the phase coordinates. 
When $N>2$, there are also situations where periodic orbits do not wind at all (as will be seen later), giving $w_a=0$ for all $a$.

However, an immediate difficulty occurs in defining winding numbers due to the appearance of quasiperiodic attractors (i.e., the torus attractor) for populations $N>2$. 
Generically, there are situations where populations are not coupled strongly enough to synchronize, and the trajectory would therefore not be closed or have a well-defined invariant. 
For example, the teal trajectories in Fig.~\ref{fig:gen_chiral}(f,h) exhibit quasiperiodicity because the third population is not synchronized with the first two, causing the trajectory to never repeat for any finite time.

As an attempt to avoid this issue, we take advantage of the property that the right-hand side of the equation of motion for the phase $\phi_a$, given by,
\begin{equation}
     \dot{\phi}_a = \sum_{b} J_{ab}\bigg(\frac{R_b}{R_a}+R_aR_b\bigg)\sin(\phi_b - \phi_a)\equiv\sum_bf_{ab}(\phi_{ab}).
     \label{eq:thisone}
\end{equation}
Eq.~\eqref{eq:thisone} is written solely in terms of the difference between the phases $\phi_{ab}=\phi_a-\phi_b$, and is independent of the total phase coordinate $\Phi=\sum_a\phi_a$ due to the $SO(2)(\subset O(2))$ symmetry (which requires the equation of motion to be invariant under the transformation $(\phi_1, \phi_2, \cdots)\rightarrow (\phi_1+\varphi, \phi_2+\varphi, \cdots)$).
The equations of motion in the $(\phi_{ab}, \Phi)$ coordinates are therefore independent of $\Phi$:
\begin{subequations}
    \begin{eqnarray}
                \dot\phi_{ab} &=& \sum_c [f_{ac}(\phi_{ac})-f_{bc}(\phi_{bc})], 
                \label{eq:phasediffOA phiab}
                \\
        \dot\Phi &=& \sum_{ab}f_{ab}(\phi_{ab}).
        \label{eq:phasediffOA Phi}
    \end{eqnarray}
\label{eq:phasediffOA}
\end{subequations}
We briefly note that, although we concentrate on the Ott-Antonsen equations Eq.~\eqref{eq:OA}, the form of the equations of motion \eqref{eq:phasediffOA} is generic to any $O(2)$ symmetric non-reciprocal system.

Because the equations for the phase differences $\phi_{ab}$ \eqref{eq:phasediffOA phiab} do not depend on the total phase $\Phi$, for the purpose of classifying the phases and phase transitions that occur in this system, we can omit the $\Phi$ equation of motion \eqref{eq:phasediffOA Phi} and solely analyze the reduced system. 
For $N=3$, the period of the limit cycles in the reduced system is generically not commensurate with the oscillations in $\Phi$. 
In this case, the $\phi_{ab}$ subsystem exhibits periodic motion with a limit cycle attractor, as shown in Fig.~\ref{fig:gen_chiral}(g,i), which plots the dynamics in these coordinates for $N=3$ populations. 
Therefore, even in the case where the trajectories in the $\phi_a$ coordinate exhibit quasiperiodic orbits (and hence Eq.~\eqref{eq:winding_a} is ill-defined),  we can safely define the winding numbers for $N=3$ populations in the $\phi_{ab}$ coordinates as,
\begin{eqnarray}
    w_{ab}=\frac{1}{2\pi}\
    \int_0^T \dot\phi_{ab}(t)dt.
    \label{eq:windingab}
\end{eqnarray}
Without loss of generality, we can choose $N-1$ of the $N(N-1)/2$ total $\phi_{ab}$-coordinates to form a coordinate system along with $\Phi$, a convenient choice being $\phi_{12}, \phi_{23},...,\phi_{N-1,N}$.
For $N>3$ populations, the $\phi_{ab}$-coordinates can still admit quasiperiodicity, which we will consider in Sec. \ref{sec:higherpop}.

A salient feature of the non-reciprocal $O(2)$ symmetric system is the existence of a one-way coupling between $\phi_{ab}$ and $\Phi$ (Eq.~\eqref{eq:phasediffOA}): $\phi_{ab}$ affects the dynamics of $\Phi$ but $\Phi$ does not affect $\phi_{ab}$. This gives rise to the following two general features.

First, the total phase $\Phi$ can persistently evolve exhibiting chiral motion, i.e.,
\begin{eqnarray}
     \avgz{\dot\Phi}\ne 0, 
\end{eqnarray}
where $\avgz{~\cdot~}$ is the average of a quantity over an attractor $z_*(t)$ for the equations of motion Eq.~\eqref{eq:OA}.
The value of $\avgz{\dot\Phi(t)}$ is crucially determined by whether or not $\mathbb{Z}_2$ symmetry is broken.
In \appref{app:z2restored}, we prove this for a generic long-time solution $z_{*}(t)$. 
For now, consider a limit cycle state that is dynamically $\mathbb{Z}_2$ restored, meaning that the $\mathbb{Z}_2$ symmetry maps the limit cycles into itself, satisfying 
\begin{eqnarray}
    z_{*}(t) \sim \bar z_{*}(t+\tau),
\end{eqnarray}
where $\sim$ again means the solutions are the same except a possible shift in the $\Phi$-coordinate. 
Because the $\mathbb{Z}_2$-symmetry transformation ($z_a\rightarrow \bar z_a$) corresponds to the transformation $(\Phi,\phi_{ab})\rightarrow (-\Phi,-\phi_{ab})$, this symmetry of the system gives the following constraint on Eq.~\eqref{eq:phasediffOA Phi}: 
\begin{equation}
\begin{aligned}
  \avgz{\dot\Phi}
  &= \avgz{\sum_{ab}f_{ab}\bigl(\phi_{ab}(t)\bigr)} \\[4pt]
  &= -\,\avgz{\sum_{ab}f_{ab}\bigl(\phi_{ab}(t+\tau)\bigr)}\\[4pt]
     &=- \avgz{\sum_{ab}f_{ab}\bigl(\phi_{ab}(t)\bigr)},
\end{aligned}
\label{eq:Phiproof}
\end{equation}
where the final equality holds because the average of an arbitrary function of the coordinates $\mathcal{O}[z(t)]$ is unaffected by any time shift $\tau$,
\begin{equation}
\begin{aligned}
  \avgz{\mathcal{O}[z(t+&\tau)]}
  = \frac{1}{T}\int_{t_0}^{t_0+T} \mathrm{d}t\;\mathcal{O}[z_*(t+\tau)] \\[4pt]
  &= \frac{1}{T}\int_{t'_0}^{t'_0+T} \mathrm{d}t\;\mathcal{O}[z(t)]
     = \avgz{\mathcal{O}[z(t)]}\,,
\end{aligned}
\end{equation}
with $t_0'=t_0+\tau$. Directly following from the equalities in Eq.~\eqref{eq:Phiproof},
$\avgz{\dot\Phi(t)}=0$ in a $\mathbb{Z}_2$-symmetric limit cycle phase.

In general, the inverse is not true: one can find examples with additional symmetries or the fine-tuning of parameters 
that exhibit spontaneous $\mathbb{Z}_2$-symmetry breaking but do not have drift in the total phase $\Phi$.
For example, reciprocal systems with spontaneous $\mathbb{Z}_2$ symmetry cannot have a dynamic steady state and therefore do not exhibit a drift of total phase $\avgz{\dot\Phi(t)}=0$.
However, for the non-reciprocal case, $\avgz{\dot\Phi(t)}$ is typically nonzero when the $\mathbb{Z}_2$ symmetry is spontaneously broken (i.e., pairs of the attractors with the opposite chirality transform to each other via the $\mathbb{Z}_2$ operation).

Second, the one-way coupling to $\Phi$ causes the emergence of the so-called critical exceptional point \cite{Fruchart2021, Hanai2020, zelleUniversalPhenomenologyCritical2024}.
The Jacobian matrix $J$ of the dynamical system \eqref{eq:phasediffOA} takes the form
\begin{eqnarray}
    J\bigl(\mathbf{x}(t)\bigr) = \left(\,
    \begin{array}{c|ccc}
        0 & \frac{\partial\dot \Phi}
        {\partial\phi_{12}} 
        &  \frac{\partial\dot \Phi}{\partial\phi_{23}} 
        &  \cdots \\[1ex]
        \hline\\[-2ex]
        0 & \frac{\partial\dot \phi_{12}}
        {\partial\phi_{12}} 
        & \frac{\partial\dot \phi_{12}}
        {\partial\phi_{23}} 
        & \cdots \\
         0 & \frac{\partial\dot \phi_{23}}
        {\partial\phi_{12}} 
        & \frac{\partial\dot \phi_{23}}
        {\partial\phi_{23}} 
        & \cdots \\
         \vdots & \vdots & \vdots & \ddots
    \end{array}
    \right)
\end{eqnarray}
where $\mathbf{x}(t) = (\Phi(t), \phi_{12}, \phi_{23},...,\phi_{N-1,N})$. 
The first column is strictly zero due to the emergence of a Nambu-Goldstone mode from the  $O(2)$ symmetry. When the attractor in the $\phi_{ab}$-dynamics is a stable fixed point ($\mathbf{x}(t)=\mathbf{x}_*$), the Jacobian takes the form (for a diagonal basis in $\phi_{a,a+1}$-space represented by $\tilde J$),
\begin{eqnarray}
    \tilde J\bigl(\mathbf{x}_*\bigr) = \left(\,
    \begin{array}{c|ccc}
        0 & A_1
        &  A_2
        &  \cdots \\[1ex]
        \hline\\[-2ex]
        0 & \mu_1
 
        & 0  
        & \cdots \\
         0 & 0 
        & \mu_2
        
        & \cdots \\
         \vdots & \vdots & \vdots & \ddots
    \end{array}
    \right)
\end{eqnarray}
where the real part of the diagonal entries are the Lyapunov exponents (i.e., $\lambda_i = {\rm Re}(\mu_i)$). The constants $A_1, A_2$ are generically non-zero when non-reciprocity is present. 
At a bifurcation with a vanishing eigenvalue ($\mu_1 = 0$), such as a pitchfork bifurcation, the first $2\times 2$ block of the Jacobian becomes a Jordan block
\begin{eqnarray}
    \tilde J\bigl(\mathbf{x}_*\bigr) = \left(\,
    \begin{array}{c|ccc}
        0 & A_1
        &  A_2
        &  \cdots \\[1ex]
        \hline\\[-2ex]
        0 & 0
 
        & 0  
        & \cdots \\
         0 & 0 
        & \mu_2
        
        & \cdots \\
         \vdots & \vdots & \vdots & \ddots
    \end{array}
    \right)
\end{eqnarray}
associated with the zero eigenvalue having algebraic multiplicity two.
This phenomenon, known as the critical exceptional point, has been shown to exhibit anomalous features (especially in the presence of stochasticity \cite{weisExceptionalPointsNonlinear2023a, Hanai2020, zelleUniversalPhenomenologyCritical2024}).

The above argument can be generalized to the case where $\phi_{ab}$ dynamics exhibit periodic orbits~\cite{weisExceptionalPointsNonlinear2023a} using the time evolution operator once around the orbit instead of $J$, as we review in \appref{app:eps}.
\subsection{\texorpdfstring{$N=2$}{N=2} population case}
\begin{figure}
    \centering
    \includegraphics[width=8cm]{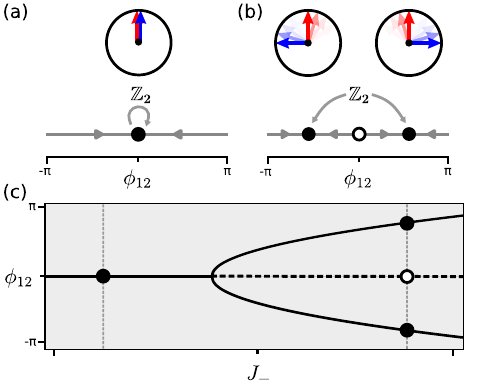}
    \caption{\strong{Two population phases and bifurcations.}
    (a-b) Two possible configurations of two (or one) stable fixed points for $N=2$ populations in the $\phi_{12}=\phi_1-\phi_2$ coordinate. The $\mathbb{Z}_2$-symmetric fixed point corresponds to the static aligned phase (a), while the $\mathbb{Z}_2$-symmetry broken fixed point corresponds to the chiral phase.
    (c) A bifurcation diagram demonstrating the pitchfork bifurcation occurring between the fixed points configurations shown in panels(a-b) where solid and dashed lines represent the phase $\phi_{12}$ of the stable and unstable fixed points versus the non-reciprocity $J_{-}=J_{12}-J_{21}$.}
    \label{fig:n2bifurcations}
\end{figure}

We begin by considering the possible phases with $N=2$ populations as studied in Ref.~\cite{Fruchart2021}. Figure~\ref{fig:n2bifurcations} shows the possible dynamics and phase transitions that may occur in the $\phi_{12}=\phi_1 - \phi_2$ subspace. Noting the $\mathbb {Z}_2 (\subset O(2))$ symmetry, unsurprisingly, the only bifurcation that occurs in the one-dimensional system is a pitchfork bifurcation in the $\phi_{12}$ equation of motion corresponding to the static-to-2-chiral transition (i.e., the non-reciprocal phase transition~\cite{Fruchart2021}). 
We emphasize once again that the $\phi_{12}$ dynamics are one-way coupled to the total phase $\Phi=\phi_1+\phi_2$ dynamics, giving rise to periodic motion in the variable $\Phi$ in the $\mathbb {Z}_2$-symmetry broken phase (i.e., the 2-chiral phase) and a critical exceptional point at the bifurcation point~\cite{weisExceptionalPointsNonlinear2023a,Fruchart2021}. 

\begin{figure}
    \centering
    \includegraphics[width=8cm]{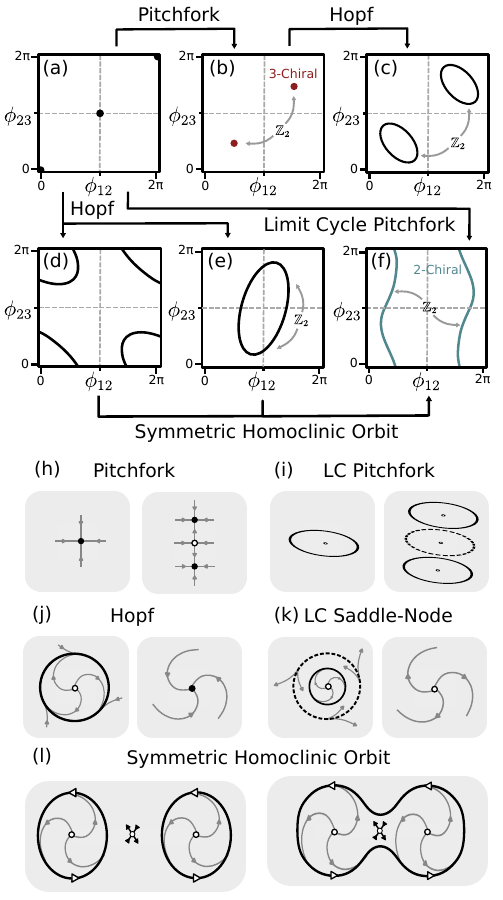}
    \caption{\strong{Three population phases and bifurcations.}
    (a-f) Possible trajectories for $N=3$ population dynamics in terms of $\phi_{12}=\phi_1-\phi_2$ and $\phi_{23}=\phi_2 - \phi_3$ coordinates.
    Here, we have restricted ourselves to cases where the winding in each direction is at most one and choosing $\phi_{23}$ to be the winding direction.
    We emphasize that the variables $\phi_{ab}$ are uni-directionally coupled to the total phase $\Phi$ due to non-reciprocity, giving rise to unique bifurcations.
    (g-l) Possible bifurcations occurring in between the trajectories shown in panels (a-f), where solid and dashed lines represent stable and unstable orbits respectively and gray arrows are typical trajectories.
    }
    \label{fig:bifurcations}
\end{figure}

\subsection{\texorpdfstring{$N=3$}{N=3} population case}
In contrast to the $N=2$ case, the $N=3$ population exhibits a diverse array of phases and phase transitions.
Figure~\ref{fig:bifurcations}(a-f)
shows the possible configurations of stable fixed points and periodic trajectories on the $(\phi_{12},\phi_{23})$ submanifold for $N=3$ population system, demonstrating that increasing the number of populations allows for far richer phase transitions.
Here, we have assumed that there exists at most one attractor per quadrant (indicated by the four different regions separated by the dashed lines in Fig. \ref{fig:bifurcations}(a-f)) of the $(\phi_{12},\phi_{23})$ plane. 
We exclude trajectories which wind in both the $\phi_{12}$ and $\phi_{23}$ directions because they can be generically eliminated through a different choice of coordinates. For example, if populations one and three are undergoing chiral motion, $\phi_{12}$ and $\phi_{23}$ will both be winding, but in the $(\phi_{13},\phi_{23})$ coordinate system, only $\phi_{23}$ will wind.
\begin{figure*}
    \centering
    \includegraphics[width=\textwidth]{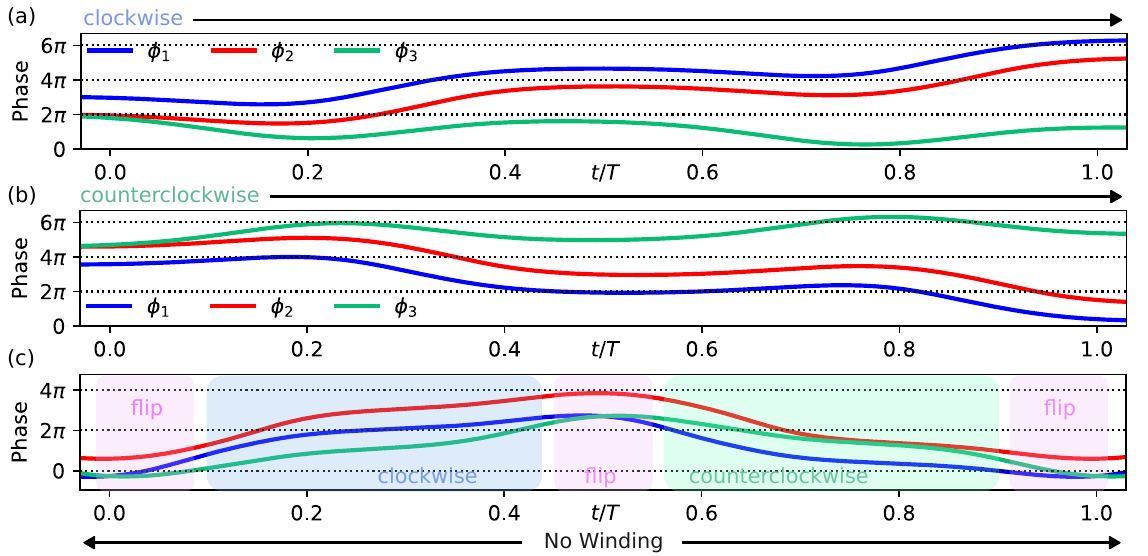}
    \caption{ \strong{Dynamics near the $\mathbb{Z}_2$-symmetric homoclinic orbit bifurcation.} 
    (a-b) Before the bifurcation, the system exhibits a chiral phase characterized by a clockwise (a) or counterclockwise (b) rotation, which are characterized by the winding number $w_{23}=1$ (a) and $w_{23}=-1$ (b), respectively. (c) After the $\mathbb{Z}_2$ SHO bifurcation takes place, the phase does not exhibit winding (i.e. $w_{12}=w_{23}=0$). We set for $(J_1,\ J_2) = (1,0.7)$ for (a) and (b) and $(J_1,\ J_2) = (1,0.5)$ for (c) with couplings $J_{ab}$ defined as $J_{12}= -0.03 + J_{1},\ J_{13}=-0.02 -J_{2},\ J_{21}= -0.01 -J_{1},\ J_{23}=0.09 + J_{2},\ J_{31}=-0.05 + J_{2},\ J_{32}=-0.01-J_{2}$, with $J_{aa}=10$ and $\Delta=0.75$. Here, $T\approx29$ is the period of the limit cycle in (c). The two chiral limit cycles are approximately half the period of the non-winding cycle away from the bifurcation.
    }
    \label{fig:z2hodynamics}
\end{figure*}
The $\mathbb{Z}_2$ symmetry manifests as $(\phi_{12},\phi_{23})\rightarrow(2\pi-\phi_{12},2\pi-\phi_{23})$, forcing a trajectory to be mapped into itself as in Figure~\ref{fig:bifurcations}(a,b,e), or to a symmetric partner as in Figure~\ref{fig:bifurcations}(b,c,f).
For a periodic orbit to map onto itself under the $\mathbb{Z}_2$ symmetry, it must be centered on a fixed point that is similarly mapped to itself, forcing the orbit to enclose either $\{(0,0),(0,\pi),(\pi,0),(\pi,\pi)\}$. The limit cycles enclosing $(0,0)$ and $(\pi,\pi)$ are respectively shown in Figure~\ref{fig:bifurcations}(d,e). The orbit that winds once in the $\phi_{12}$ or $\phi_{23}$ direction generically has a reflected partner, as it would otherwise be required to pass through one of the fixed points listed above.

\subsubsection{Phase transitions analogous to \texorpdfstring{$N=2$}{N=2} population case}

Given the possible trajectories, we can now examine the types of bifurcations that may occur in the $N=3$ population system.
The pitchfork bifurcation illustrated in Fig.~\ref{fig:bifurcations}(h) occurs between the configuration of fixed points in panels (a-b). 
Here, the symmetric fixed point that corresponds to all populations aligned ($\phi_{ab}=0$) or anti-aligned ($\phi_{ab}=\pi$) will have the $\Phi$-coordinate constant in time, and the symmetry broken pair of fixed points (with $\phi_{ab}\ne 0,\pi$) will have the $\Phi$ coordinate oscillating at $\pm\omega$.
In the original $\{\phi_a\}$ coordinates, the populations all exhibit a chiral motion characterized by non-trivial winding numbers $w_1,w_2,w_3=\pm1$, 
corresponding to the $3$-chiral phase.
We note that the pitchfork bifurcation can also occur on the $\phi_{ab}$ manifold for arbitrary $N$, still coinciding with a critical exceptional point as demonstrated in \appref{app:eps} for $N=3,4,5$. This is not surprising, as the pitchfork bifurcation is the most generic codimension‑one bifurcation in systems with an underlying $\mathbb{Z}_2$ symmetry.

Next, consider the limit where the third population is decoupled from the rest $J_{23}=J_{32}=0$. The first two populations may exhibit a non-reciprocal phase transition from the static phase corresponding to a line of fixed points with $\phi_{12}=0,\pi$ and arbitrary $\phi_{23}$ to the $2$-chiral phase where $\phi_1 = \Omega t, \phi_2 = \Omega t + \phi_{12}$ (with $\phi_{12}={\rm const.}$),
while the phase $\phi_3$ remains constant.
The static-to-2-chiral bifurcation (Fig.~\ref{fig:bifurcations}(a) transitioning to (f)) corresponds to a limit cycle pitchfork in the $\phi_{ab}$ coordinates as drawn in Fig.~\ref{fig:bifurcations}(i).  This phase transition disappears with any finite coupling to $\phi_3$ therefore requiring fine tuning of multiple parameters to restore. The additional coupling breaks the symmetry that creates the line of static fixed points causing the occurrence of additional intermediate transitions. 
\subsubsection{Novel Phase Transitions for \texorpdfstring{$N=3$}{N=3} populations}

More exotic phase transitions with no counterpart in the $N=2$ population case also occur for three populations.
An interesting example is the  $\mathbb{Z}_2$-Symmetric Homoclinic Orbit (SHO) bifurcation which merges two $\mathbb{Z}_2$-broken orbits 
(Fig.~\ref{fig:bifurcations}(d,e)) 
into a single stable orbit. 
The simplest form of the bifurcation (illustrated in the space $\mathbb{R}^2$ for simplicity) is shown in Fig.~\ref{fig:bifurcations}(l).
Before the bifurcation, two limit cycles exist. At the bifurcation, the two limit cycles become homoclinic orbits which exit the saddle point along the unstable manifold and re-enters the same saddle along the stable manifold. After the bifurcation, the two cycles are merged into a single periodic orbit.

The SHO bifurcation in the multipopulation system changes the winding number of the limit cycles on the torus (Fig.~\ref{fig:bifurcations}(d-f)). The trajectories in the phase variables $\phi_a$ on either side of the phase transition are shown in Fig. \ref{fig:z2hodynamics}.
Before the bifurcation, two 2-chiral trajectories wind with $w_{23}=\pm1$. 
After the bifurcation, the two 2-chiral orbits are glued together at a point to form one orbit with zero net winding. The new trajectory does so by alternating between the CW and CCW trajectories of the chiral limit cycles in each oscillation.
The $\mathbb{Z}_2$ conjugation symmetry as described above is broken on both sides of the SHO phase transition. In the 2-chiral phase, the $\mathbb{Z}_2$ conjugation symmetry is spontaneously broken. In the non-winding phase, a new $\mathbb{Z}_2$ symmetry is \emph{dynamically restored} because the cycle is reflected into itself. The  nontrivial dynamical $\mathbb{Z}_2$ symmetry manifests as $\bar{z}_a(t) \sim z_a(t+T/2)$. The symmetry must hold by the observation that the conjugation operation maps the limit cycle into itself and is not the identity map; because the $\mathbb{Z}_2$ is an involution, $\bar z_a(t)$ can only be the original trajectory time shifted by $T/2$ or $0$. The emergence of chiral cycles from the $\mathbb{Z}_2$ restored limit cycle serves as an exotic form of symmetry breaking. Near the transition, the limit cycle period diverges as the bifurcation is approached. The divergence of the order parameter and the period are further explored in \appref{app:3pop}.
\begin{figure*}
    \includegraphics[width=\textwidth]{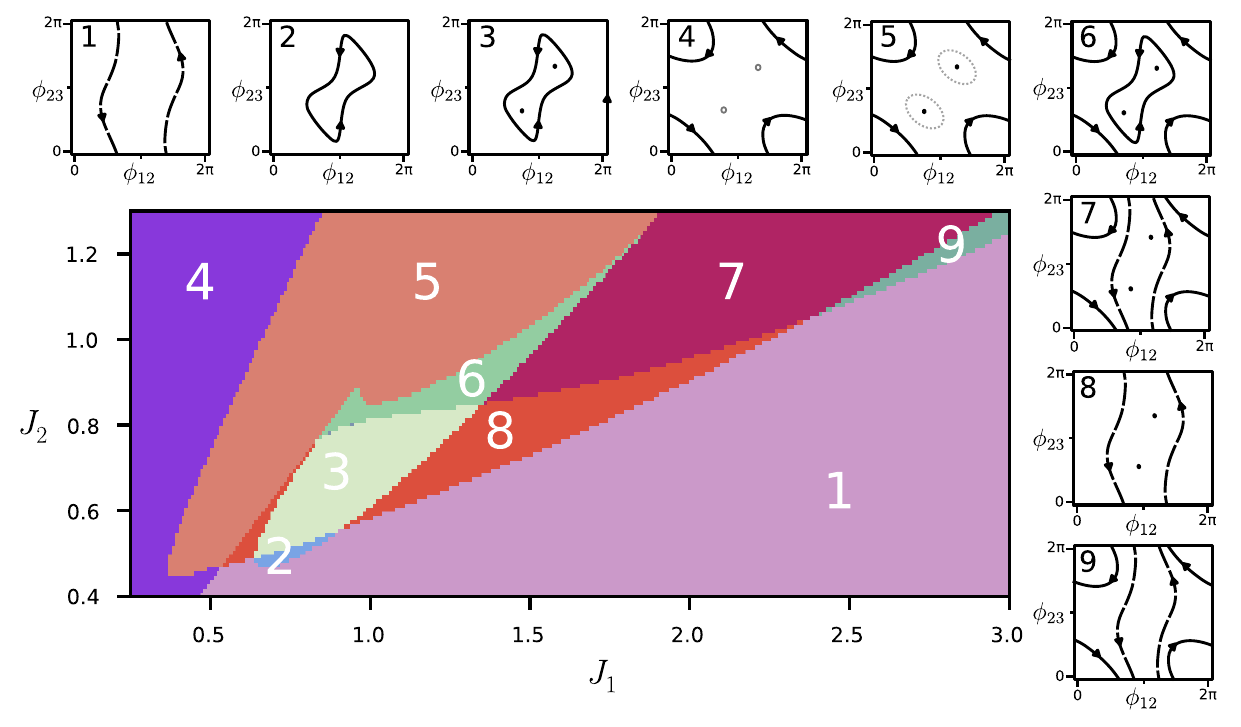}
    \caption{\strong{Numerical phases for three populations in the $\phi_{a,a+1}=\phi_a-\phi_{a+1}$ coordinate.} A slice of the phase space of the 3-population model with parameters 
    $J_{12}= -0.03 + J_{1},\ J_{13}=-0.02 -J_{2},\ J_{21}= -0.01 -J_{1},\ J_{23}=0.09 + J_{2},\ J_{31}=-0.05 + J_{2},\ J_{32}=-0.01-J_{2}$, with $J_{aa}=10$ and $\Delta=0.75$ (identically to Fig. \ref{fig:z2hodynamics}), where the couplings were shifted randomly to break symmetry. The configuration of stable solutions in each phase is given by the plot with the corresponding number. Notice that Phase 8 is split into two disconnected components by Phase 3. The $N=3$ chiral state appears in the plot as a single dot, limit cycles with no winding are solid lines, and 2-chiral limit cycles are represented with dashed lines. Examples of unstable trajectories that exist are shown in gray for Phases 4 and 5. The 2-chiral and 3-chiral trajectories are also plotted in Fig.} \ref{fig:gen_chiral}(f).
    \label{fig:n3phasediagram}
\end{figure*}
Unlike many phase transitions, the mean-field dynamics is undergoing a \emph{global} bifurcation. As the name suggests, the bifurcation cannot be understood from only a small neighborhood of a fixed point or limit cycles ~\cite{Kuznetsov2004}, as is typically the case with bifurcations which change the linear stability of an attractor. Instead, global bifurcations require analyzing the global topology of the flow lines of the equations of motion. 
It is worth noting that the homoclinic orbit bifurcation occurs without an additional zero mode emerging at the bifurcation (i.e., an additional Floquet multiplier approaching unity at the bifurcation, closing the gap to the first excited mode), even though the limit cycle period is diverging. This can be most easily analyzed in the homoclinic orbit bifurcation normal form ~\cite{Kuznetsov2004}. 
    

Two additional bifurcations can occur in the $N=3$ population model: The Hopf bifurcation (Fig. \ref{fig:bifurcations}(j)) and the {Limit Cycle Saddle Node (LCSN) bifurcation (Fig. \ref{fig:bifurcations}(k)). In the supercritical Hopf bifurcation, a stable periodic orbit without winding (such as (b) and (e)) shrinks into a stable fixed point.
In the case of the subcritical Hopf bifurcation, the unstable limit cycle forms the domain of attraction, which shrinks to zero as the attractive fixed point becomes unstable.
Finally, the LCSN bifurcation creates or destroys stable/unstable limit cycle pairs which can occur for any type of periodic orbit.

\subsubsection{Numerical Confirmation of Limit Cycle Phases}

In Fig. \ref{fig:n3phasediagram}, we demonstrate all the above phase transitions in a portion of the phase diagram for the three-population model by interpolating between the $2-$chiral regime where there is weak couplings to the third population and the strong non-reciprocal coupling three-population regime. 
This portion of the phase diagram contains nearly all combinations of the (b,d,e,f) trajectories of Fig. \ref{fig:bifurcations}.
We find that the existing periodic orbits are not strongly dependent on how the couplings are directed between the populations at strong non-reciprocity. See \appref{app:3pop} for the same phases on an alternative coupling network.
We note that, in constructing the phase diagram in Fig.~\ref{fig:n3phasediagram}, it was practically easier to compute the phase current as defined in \appref{app:3pop} (which is averaged over the trajectory for long times) instead of directly computing the winding number (which requires a parametrization of the trajectory over one period). The evaluation of the phase current on each stable periodic orbit at each point in parameter space produces the phase diagram shown in Fig. \ref{fig:n3phasediagram}. 
Each line in the phase diagram in Fig. \ref{fig:n3phasediagram} represents one of three bifurcations shown Fig. \ref{fig:bifurcations}(j,k,l) occurring on one of the existing periodic orbits. For example, the limit cycle saddle-node bifurcation occurs between Phases 5 and 6 and the SHO bifurcation occurs between Phases 8 and 3. The (c) trajectories are absent from the phase plots because they only appear in this region of parameter space as unstable limit cycles forming the domain of attraction around the 3-chiral trajectory as shown in Fig. \ref{fig:n3phasediagram} Panel 5. The subcritical Hopf bifurcation destabilizes the $3$-chiral trajectory between Phases 5 to 4. Supercritical Hopf bifurcations which stabilize the static phase can also occur in other portions of parameter space as shown in \appref{app:3pop}. Notice the lack of co-occurrence of (e,f) trajectories in the phases in Fig. \ref{fig:n3phasediagram}
for a simple reason: the contractible (e) trajectory only appears via the SHO bifurcation of the 2-chiral (f) trajectories.

\section{Phase Transitions in Systems with \texorpdfstring{$N\geq4$}{N>3} Populations}
\label{sec:higherpop}
\begin{figure*}
    \centering
    \includegraphics[width=\textwidth]{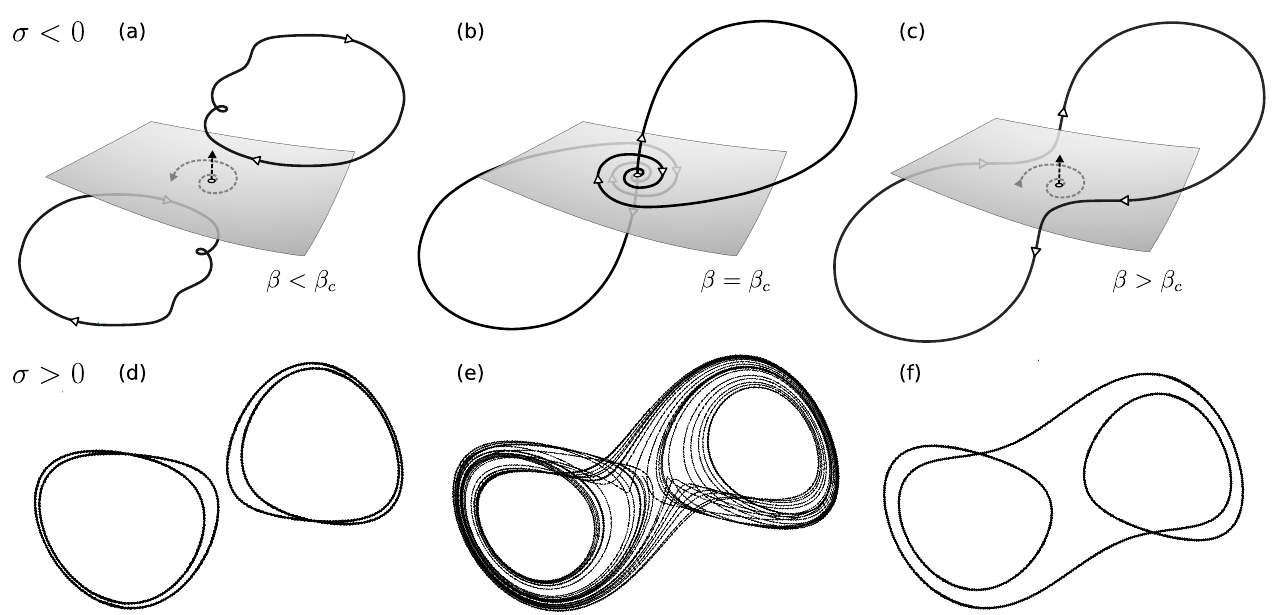}
    \caption{\strong{Saddle-Focus Homoclinic Orbit Bifurcation in $\mathbb{R}^3$.}
    The dynamics near the symmetric saddle-focus homoclinic orbit bifurcation for a negative(a-c) and positive(d-f) saddle quantity $\sigma = \Real(\lambda_{1}) + \Real(\lambda_{2,3})$, the sum of the saddle-focus's leading stable and unstable eigenvalue. Here, the bifurcation is drawn in $\mathbb{R}^3$, but occurs analogously with the limit cycles wrapping around $\mathbb{T}^3$ in the $\phi_{a,a+1}$ coordinates.
    In (a-c), a saddle-focus is shown in the center with a one dimensional unstable manifold (the black dashed line) and with a two dimensional stable manifold corresponding to the pair of complex eigenvalues with negative real part (gray surface). Before the bifurcation, two limit cycles approach the saddle-focus (a). As a bifurcation parameter $\beta$ is tuned to the bifurcation point $\beta_c$, the two limit cycles become homoclinic orbits that connect the stable and unstable manifolds of the saddle (b). After the bifurcation, a single limit cycle oscillates around the saddle. In (d-f), we demonstrate the behavior of a system near a the $\mathbb{Z}_2$-symmetric saddle-focus homoclinic orbit bifurcation (in $\mathbb{R}^3$) with saddle quantity $\sigma>0$, which exhibits a proliferation of saddle limit cycles resulting in chaotic behavior. 
    To visualize the trajectories, we used the Chua model (Eq. C.7.1 in~\cite{shilnikovMethodsQualitativeTheory2001a}) with parameters $b=4.25,$ $ a= 3.5$ in (d),  $b=4,$ $ a= 3.5$ in (e), and $b=3.85,$ $ a= 3.5$ in (f).
    }
    \label{fig:sf_homo}
\end{figure*}
Thus far, we have systematically found the possible phases and phase transitions occurring for the $N=3$ population case.
As the number of populations increases beyond three, the range of possible dynamical attractors becomes richer and more complex, which not only includes limit cycles with various winding numbers, but also quasiperiodic and chaotic attractors. 
We can now apply the framework introduced above for $N=3$ populations to study the bifurcations of periodic and quasiperiodic orbits for the case of $N\ge 4$ populations. 

All of the bifurcations discussed in Fig. \ref{fig:bifurcations} can occur to transition between the various possible periodic orbits on the $\phi_{ab}$ manifold. 
As is well known in nonlinear dynamics, bifurcations in low dimensional systems can similarly occur in a higher dimensional system through a reduction to the center manifold, making the three population bifurcations possible in larger systems.
That being said, there are one-parameter bifurcations that can only occur for $N>3$ which we explore in the following section.

\subsection{Higher Population Homoclinic Orbit Bifurcations}

An example of a bifurcation of periodic orbits in the $\phi_{a,a+1}$ coordinate (which can become quasiperiodic orbits in the $\phi_a$ coordinates due to incommensurate oscillations with total phase $\Phi$),
that can emerge uniquely for $N>3$ population systems,
is schematically shown in Fig.~\ref{fig:sf_homo}.
Note that, for simplicity, we have embedded the orbits 
in $\mathbb{R}^3$, but the bifurcation occurs similarly in the $\phi_{a,a+1}$ coordinates with the limit cycles wrapped around the $\mathbb{T}^3$ phase space manifold for $N=4$ populations.
As illustrated here, 
the $\mathbb{Z}_2$-symmetric homoclinic orbit bifurcation explored for three populations in Figs. \ref{fig:bifurcations} and \ref{fig:z2hodynamics} can exhibit alternative forms for $N>3$. For $N=3$ populations, the eigenvalues $\lambda_i$ of the Jacobian matrix on the  $(\phi_{12},\phi_{23})$ manifold at the saddle point must be purely real, with $\lambda_1>0,\lambda_2<0$. 
In higher dimensions, a metastable point could also correspond to a saddle, saddle-focus, or focus-focus depending on whether the leading stable/unstable eigenvalues are complex conjugate pairs. Generic homoclinic‐orbit bifurcations under appropriate transversality and non-resonance conditions can be reduced by center manifold reduction to a two, three, or four-dimensional system corresponding to which of the metastable fixed points above occur~\cite{shilnikovMethodsQualitativeTheory1998, Kuznetsov2004, sandstedeCenterManifoldsHomoclinic2000, homburgGlobalAspectsHomoclinic1996, turaevBifurcationsDynamicalSystems1991}. 

Figure~\ref{fig:sf_homo}(a-c) shows an example that can occur for $N=4$ populations:
the saddle-focus case of the $\mathbb{Z}_2$-symmetric homoclinic orbit bifurcation~\cite{xingOrderedIntricacyShilnikov2021}.
Here, we assume that the complex conjugate pair of eigenvalues of the saddle-focus have $\Real(\lambda_{2,3})<0$ and the remaining eigenvalue satisfies
$\lambda_1>0$. 
Two limit cycles meet the saddle-focus at the bifurcation and become homoclinic orbits which connect the fixed point's stable and unstable manifolds. After the bifurcation, a single merged limit cycle persists. In the dynamics of Eq.~\eqref{eq:OA},
any symmetric homoclinic orbit bifurcation for $N$-populations generically merges two chiral limit cycles into a single limit cycle with no winding.

While the saddle-focus bifurcation as shown in Fig. \ref{fig:sf_homo}(a-c) proceeds analogously to the saddle case shown in Fig. \ref{fig:bifurcations}(l), Shilnikov's theorems~\cite{Shilnikov1965,Shilnikov1967} established that near the saddle-focus homoclinic orbit, a chaotic or quasi-chaotic attractor~\cite{ovsyannikovSystemsHomoclinicCurve1991, barenblattNonlinearDynamicsTurbulence1983} can be produced. Define the saddle quantity $\sigma$ for the saddle-focus to be $\sigma  = \Real(\lambda_{1}) + \Real(\lambda_{2,3})$. For $\sigma<0$, the dynamics near the $\mathbb{Z}_2$-symmetric bifurcation generically proceed as shown in Fig. \ref{fig:sf_homo}(a-c). However, Shilnikov proved~\cite{shilnikovMethodsQualitativeTheory1998,Kuznetsov2004} that near a saddle-focus homoclinic orbit bifurcation under generic conditions and for $\sigma>0$, there are an infinite number of saddle limit cycles in a neighborhood of the homoclinic orbit sufficiently close to the bifurcation. The cycles appear through intermediate bifurcations that result in the emergence of chaos~\cite{gaspardWhatCanWe1983}. We demonstrate the chaotic dynamics near the homoclinic orbit bifurcation in a simplified model in Fig. \ref{fig:sf_homo}(d-f), where far away from the SHO bifurcation, a single limit cycle or limit cycle pair exists just as in the $\sigma < 0$ case.
The focus–focus homoclinic orbit bifurcation also exhibits a similar path to chaos~\cite{fowlerBifocalHomoclinicOrbits1991, Kuznetsov2004}, occurring in at least $N=5$ populations in our system.
\subsection{Quasiperiodicity}
\begin{figure*}
    \centering
    \includegraphics[width=\textwidth]{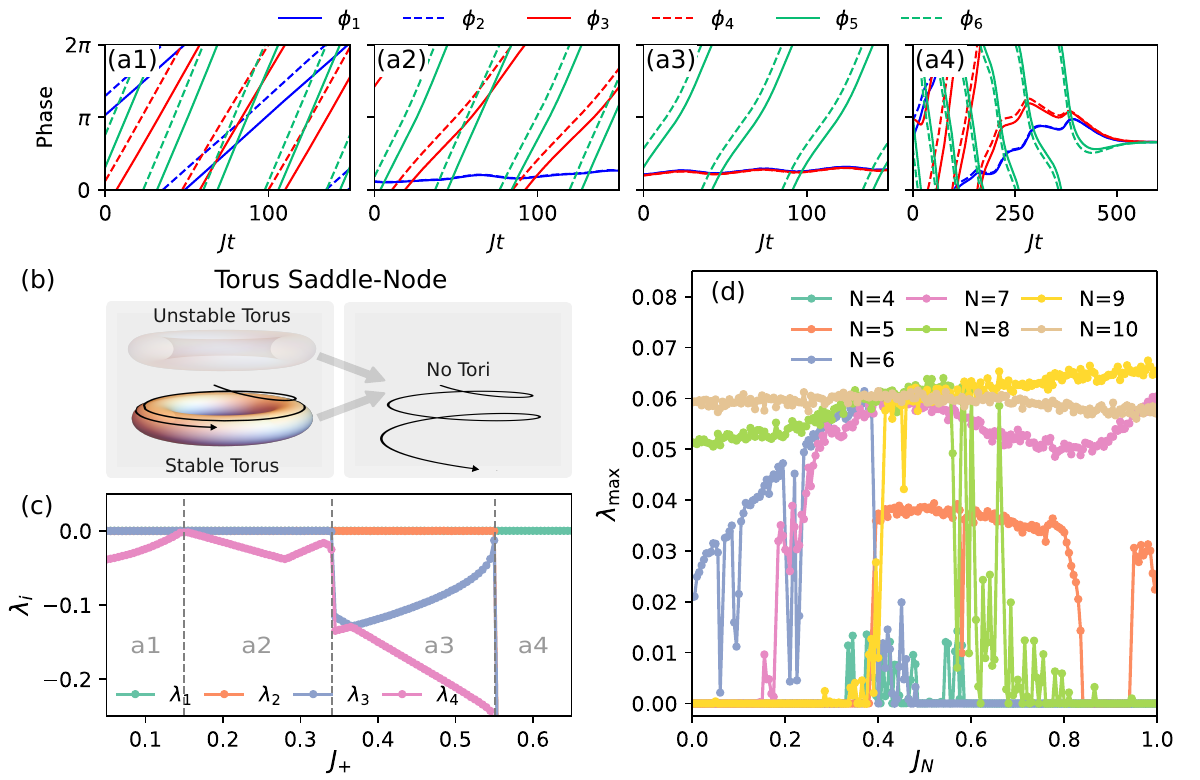}
    \caption{\strong{Quasiperiodicity and chaos.} (a1-a4) The phases $\phi_a$ over non-dimensionalized time $Jt$ where $J=J_{aa}=10$ for the four different dynamical phases existing between the aligned phase (a4) where the steady state phase is constant and equal for all populations, and the $3$-torus Phase (a1) where the three pairs of populations oscillate with incommensurate frequencies. For (a1-a4), $J_+= 0.1,0.25,0.4,0.6$ respectively. The parameters are chosen as $\Delta_a = 1.5$, $J_{ab}=10\mathbb{I}_{ab}+\frac{1}{N}\bigl(J_+(\mathcal{S}_{ab} +\mathcal{N}_{ab}) +  \mathcal{A}^{q}_{ab}\bigr)$, where $\mathcal{A}^{q}_{ab}=\bigl(\delta_{a2}\delta_{b1}-\delta_{a1}\delta_{b2}\bigr)
          +1.5\bigl(\delta_{a3}\delta_{b4}-\delta_{a4}\delta_{b3}\bigr)
          +2\bigl(\delta_{a5}\delta_{b6}-\delta_{a6}\delta_{b5}\bigr)$,     
    $\mathcal{S}_{ab}=1$ if $a\neq b$ and zero otherwise, $\mathcal{N}_{ab}=0$ if $a=b$ and otherwise $\mathcal{N}_{ab} = \mathcal{N}_{ba}\sim\mathcal{N}(0,0.05^2)$. (b) A depiction of the torus saddle-node bifurcation. On the left, a stable and unstable torus exists with nearby initial conditions converging to the stable torus attractor. On the right, the stable and unstable tori annihilate and the trajectory falls to another region of phase space after the bifurcation. (c) A plot of the three maximum Lyapunov exponents for three pairs ($N=6$) of populations through the four regimes shown in (a1-a4). Dashed lines indicate a  bifurcation where the dynamics of subsets of populations phase lock with each other. (d) A plot of the maximum Lyapunov exponent while varying the non-reciprocal coupling from the first $N-1$ populations to the $N^\text{th}$.
    The parameters are chosen as $\Delta=0.5$, $J_{ab}=10\mathbb{I}_{ab}+\frac{1}{N}(\mathcal{A}_{ab} + \delta_{b,N}(1-J_N) - \delta_{N,a}(1-J_N)+\mathcal{N}_{ab})$, where $\mathcal{A}_{ij}= \operatorname{sgn}(i-j)$ and $\mathcal{N}_{ab}$ is Gaussian distributed as $\sim\mathcal{N}(0,0.25^2)$.
    }
    \label{fig:les}
\end{figure*}

The winding numbers as defined in \eqref{eq:windingab} remain the appropriate topological numbers to characterize the attractive solution for a fixed point or limit cycle for an arbitrary number of populations in the $\phi_{ab}$ coordinates. Quasiperiodic dynamics can occur in the $\phi_{ab}$ coordinates when there are more than two synchronized subsets of populations. As a concrete example, consider a scenario involving three subsets of populations, each subset consisting of exactly two non-reciprocally coupled populations. Specifically, let populations $a=1,2$ form one subset, 
$a=3,4$ form a second subset, and 
$a=5,6$ form the third subset. Within each subset, the pair of populations are coupled non-reciprocally causing each pair to oscillate internally with a distinct frequency. When the three frequencies associated with these three non-reciprocally coupled pairs are mutually incommensurate (Fig. \ref{fig:les}(a1)), the combined dynamics of all six populations become quasiperiodic. Even when the total phase $\Phi$ is projected out, two incommensurate frequencies on the $\phi_{a,a+1}$ coordinate manifold still remain. 
For each of the three synchronized pairs of oscillators, there is a corresponding zero Lyapunov exponent as is seen in the leftmost portion of Fig. \ref{fig:les}(c) (where we obtained $\lambda_1=\lambda_2=\lambda_3=0$).
The three zero Lyapunov exponents in this phase would imply the attractor is a 3-torus, the $3D$ analogue of the $2D$ torus or ``donut" surfaces seen in Fig. \ref{fig:gen_chiral}.
The 3-torus attractor is embedded into an ($N=6$)-torus phase space. 

Formally, one can assign winding numbers to such quasiperiodic attractors as follows.
Consider a quasiperiodic attractor with $M$ zero Lyapunov exponents, one being due to the total phase $\Phi$. In $\phi_{ab}$ coordinates, there is an $(M-1)$-torus quasiperiodic attractor embedded into the $(N-1)$-torus phase space spanned by the coordinates $\{\phi_{12},\dots,\phi_{N-1,N}\}$. The $(M-1)$-torus can be arbitrarily parametrized by the coordinates $(\theta_{1},\dots,\theta_{M-1})$. 
Let $\varphi:\mathbb{T}^{\,M-1}\to\mathbb{T}^{\,N-1}$ be the embedding map $\varphi(\theta_{1},\dots,\theta_{M-1})=
\bigl(\phi^*_{12}(\vec\theta),\dots,\phi^*_{N-1,N}(\vec\theta)\bigr)$ of the quasiperiodic attractor's surface with $\theta_a \in [0, 2\pi)$. Here, each component $\varphi_a = \phi^*_{a,a+1}$ of the map $\varphi$ corresponds to a coordinate of the $\phi_{a,a+1}$ manifold.
We can associate to the embedding $\varphi$ a winding matrix with $W\in\mathbb{Z}^{(M-1)\times(N-1)}$ and

\begin{equation}
W_{ma}\;=\;
\frac{1}{2\pi}
\int_{0}^{2\pi}
\frac{\partial \varphi_{a}}{\partial \theta_{m}}
\bigl(\theta_{1},\dots,\theta_{m-1},t,\theta_{m+1},\dots,\theta_{M-1}\bigr)\,
dt,
\label{eq:wmat}
\end{equation}
where each entry of $W_{ma}$ is an integer describing how many times the $m$th cycle of $\mathbb{T}^{M-1}$ wraps around the $a$th cycle of $\mathbb{T}^{N-1}$. In the case of a limit cycle, the winding numbers in Eq.~\eqref{eq:windingab} are recovered when $\varphi_a(\theta)$ is chosen to be the natural time parametrization $\phi_{a,a+1}(t)$. The winding matrix $W$ is again fixed under ambient isotopies of $\varphi$.

As an example, consider the Phase (a1) depicted in the leftmost part of Fig. \ref{fig:les}(c) before the first dashed line,
which has eight quasiperiodic attractors
on the $\phi_{ab}$ manifold corresponding to the different combinations of oscillation chirality for each pair of populations. 
In the limit where the pairs are decoupled, the quasiperiodic attractors can be parameterized in the $\phi_{a,a+1}$ coordinates as 
\begin{equation}
    \varphi = (0,\epsilon_1\theta_1,0,\epsilon_2 \theta_2,0),
    \label{eq:parametrize}
\end{equation}
where $\varepsilon_1,\varepsilon_2\in \{-1,1\}$ are chosen such that the direction of the parametrization respects the net direction of the dynamics, which can be different for the different attractors similar to the limit cycle case. 

From the parametrization in Eq. \eqref{eq:parametrize}, the winding matrix (Eq.~\eqref{eq:wmat}) is given by 
\begin{eqnarray}
    W_{ma} = \varepsilon_{1}\,\delta_{m1}\delta_{a2} + \varepsilon_{2}\,\delta_{m2}\delta_{a4}.
\end{eqnarray} 
The matrix captures how the first ($m=1$) coordinate of the torus is wrapped around $\phi_{23}$ ($a=2$) and the second ($m=2$) coordinate of the torus is wrapped around $\phi_{45}$ ($a=4$). The components $W_{ma}$ remain the same until a bifurcation occurs, which can change the topology of the attractor.

We note that the parameterization of the torus $\varphi$ is not unique, and in general affects the winding matrix $W$.
To fix this redundancy, we consider the following procedure.
Due to the standard result that the mapping class group of $\mathbb{T}^n$ is $\mathrm{GL}_n\!\bigl(\mathbb{Z}\bigr)$
 ~\cite{farbPrimerMappingClass2012, birmanBraidsLinksMapping2016}, 
it directly follows that a full row rank winding matrix $W$ can be transformed into the row Hermite normal form $H$ such that $H=UW$ for some square unimodular matrix $U$, which is unique ~\cite{schrijverTheoryLinearInteger2011, arnoldAbelianGroupsRepresentations2000} and therefore removes the parametrization redundancy. The Hermite normal form generalizes the reduced row echelon form to integer matrices, still satisfying $H_{ab}=0$ for $b<a$.
As a minor remark in the case of Phase (a1) of the $N=6$ population system, the choice of parametrization \eqref{eq:parametrize} already resulted in the winding matrix being in the unique Hermite normal form up to the choice of signs.


Because the winding matrix will only change at a bifurcation, it can be used to classify phases with quasiperiodic attractors in terms of their topological numbers.
The simplest bifurcation of a torus attractor is a torus saddle-node bifurcation (Fig. \ref{fig:les}(b)), where a stable and an unstable torus coalesce and annihilate analogous to the limit-cycle saddle-node shown in Fig. \ref{fig:bifurcations}(k). This occurs at the middle dashed line in Fig. \ref{fig:les}(c) between phases (a2) and (a3). From left to right, the attractor corresponding to the dynamics in each phase (in the $\phi_a$ coordinates) is $3$-torus (a1), $2$-torus (a2), limit cycle (a3), fixed point (a4) corresponding to the static aligned phase. A torus pitchfork bifurcation occurs from (a1) to (a2), a torus saddle-node bifurcation from (a2) to (a3), and a limit cycle pitchfork bifurcation from (a3) to (a4). After the torus saddle-node bifurcation between phases (a2) and (a3), the steady state attractor for the dynamics is a limit cycle with different winding numbers $w_{a,a+1}=(0,0,0,\varepsilon,0)$.
In the absence of chaos, a generic phase of the $N$-population model will consist of some combination of multistable quasiperiodic tori, limit cycles, and fixed points, each classifiable by its topological winding numbers; however, the total number of distinct phases in models with many populations grows large from the many distinct possible combinations of attractors.
Between any two distinct phases, a bifurcation or sequence of bifurcations occurs, which may be one already discussed such as those in Fig. \ref{fig:bifurcations} or an additional bifurcation of quasiperiodic attractors such as a phase locking fold bifurcation or a Neimark–Sacker bifurcation~\cite{Kuznetsov2004}.

\subsection{Chaos}
Chaotic attractors cannot be as simply described with the symmetry and topology arguments used thus far. Fortunately, for large self coupling, chaos is absent in the three population model since only two degrees of freedom dominate the system dynamics, $\phi_{12}$ and $\phi_{23}$. Chaos requires at least three degrees of freedom to occur, and our numerical simulations of the three-population model revealed no chaotic behavior.

For $N\geq4$, chaotic dynamics become increasingly prevalent near the non-reciprocal limit. In Fig. \ref{fig:les}(d), we plot the maximal Lyapunov exponent for a system that begins near the limit of $N-1$ non-reciprocally coupled populations and varies the coupling from the first $N-1$ populations to the $N^\text{th}$. See the figure caption for the precise parameters used.
From four to ten populations, much of this parameter regime becomes chaotic near this non-reciprocal limit. Periodic behavior persists in other regions of parameter space, either at slightly larger $\Delta$, larger self coupling, smaller non-reciprocal coupling, or smaller variance in the couplings.

\section{Discussion}
Non-reciprocal systems exhibit novel bifurcations that are not allowed in equilibrium systems, opening a landscape of new phase transitions and statistical phenomena. The statistical physics of many non-reciprocal populations admits arbitrarily complex phases with multistable limit cycles and quasiperiodic attractors of various topologies, and with sufficiently strong non-reciprocity, even chaotic attractors. We have explored a framework for understanding the possible phases and phase transitions that can occur for an arbitrary number of populations. In particular, we have explored the $N$-chiral phase emerging as a distinct dynamical behavior for $N$ populations. Using the three-population model as a representative example, we systematically identified the limit-cycle phases that exist between the 2‑chiral and 3‑chiral phases, as well as exotic phase transitions such as the $\mathbb{Z}_2$-symmetry-restoring homoclinic orbit bifurcation. Generic systems offer multiple interesting paths to $\mathbb{Z}_2$-symmetry restoring phase transitions through the various Shilnikov homoclinic orbit bifurcations. For larger numbers of populations, we introduce a classification for all non-chaotic attractors in terms of their phase space topology. 

Within the $N$-chiral phase for larger collections of populations, there are also transitions between different numbers of out-of-phase synchronized groups, which has been explored in ecology~\cite{vandermeerNewFormsStructure, gironSynchronizationUnveilsOrganization2016a}. In 
\appref{app:npop}, we demonstrate such transitions between numbers of synchrony groups that behave similarly to a crossover phenomena for $\Delta>0$. Because the states with different numbers of "synchrony groups" have limit cycles with the same topology, it could be suspected that these transitions would not be true phase transitions.

Arbitrary bifurcations begin to appear in the mean-field dynamics in the presence of multiple populations of agents, suggesting any known bifurcation will conceivably occur as a phase transition in a model with a sufficient number of non-reciprocally interacting populations. Extending the bifurcations in this work to fluctuating spatially extended systems could correspond to novel critical phenomena beyond a mean field approximation, as well as possibly having interesting effects on the topological defects of the theory.
Non-reciprocity can also give rise to interesting phenomena such as order-by-disorder phenomena~\cite{Fruchart2021,hanaiNonreciprocalFrustrationTime2024}. Further study of such effects in a many-component system, where multiple dynamic attractors can be stabilized, is an additional interesting avenue of research.

The statistical mechanics of emergent non-reciprocally interacting systems is becoming a multidisciplinary tool to understand the behavior of many systems, ranging from neuroscience and ecological systems to open quantum systems. Exploring how these rich phenomena manifest across various non-equilibrium contexts remains an exciting and promising path for future studies, and could reveal unifying principles that underlie non-reciprocal interactions across different physical systems.
\\

\begin{acknowledgments}
{\it Acknowledgments.---} 
We acknowledge valuable discussions and input on the manuscript from P. Littlewood. We thank S. Colt and J. Salisbury for a close reading.
RH was supported by a Grant in Aid for
Transformative Research Areas
(No. 25H01364),
for Scientific Research (B) (General)
(No. 25K00935),
and for Research Activity Start-up from JSPS in Japan (No. 23K19034) and the National Research Foundation (NRF) funded by the Ministry of Science of Korea (Grant No. RS-2023-00249900).
CW was partially supported by NSF-MPS-PHY award 2207383.
This research benefited from Physics Frontier Center for Living Systems funded by the National Science Foundation (PHY-  2317138).
\end{acknowledgments}
\bibliographystyle{apsrev4-2}
\bibliography{biblio, extra}

@article{Ginelli2013,
 doi = {10.1088/1751-8113/46/25/254005},
 title = {Covariant {L}yapunov vectors},
 author = {Francesco Ginelli and Hugues Chat{\'{e}} and Roberto Livi and Antonio Politi},
 journal = {Journal of Physics A: Mathematical and Theoretical},
 volume = {46},
 number = {25},
 pages = {254005},
 year = {2013},
 month = {jun},
 publisher = {{IOP} Publishing}
}

@article{Fruchart2021,
 doi = {10.1038/s41586-021-03375-9},
 title = {Non-reciprocal phase transitions},
 author = {Michel Fruchart and Ryo Hanai and Peter B. Littlewood and Vincenzo Vitelli},
 journal = {Nature},
 volume = {592},
 number = {7854},
 pages = {363--369},
 year = {2021},
 month = {apr},
 publisher = {Springer Science and Business Media {LLC}}
}

@article{Saha2020,
 doi = {10.1103/physrevx.10.041009},
 title = {Scalar Active Mixtures: The Nonreciprocal {C}ahn-{H}illiard Model},
 author = {Suropriya Saha and Jaime Agudo-Canalejo and Ramin Golestanian},
 journal = {Physical Review X},
 volume = {10},
 number = {4},
 pages = {041009},
 year = {2020},
 month = {oct},
 publisher = {American Physical Society ({APS})}
}

@book{Kuznetsov2004,
 doi = {10.1007/978-1-4757-3978-7},
 title = {Elements of Applied Bifurcation Theory},
 author = {Yuri A. Kuznetsov},
 year = {2004},
 publisher = {Springer New York}
}

@article{Hanai2019,
 doi = {10.1103/physrevlett.122.185301},
 title = {Non-{H}ermitian Phase Transition from a Polariton {B}ose-{E}instein Condensate to a Photon Laser},
 author = {Ryo Hanai and Alexander Edelman and Yoji Ohashi and Peter B. Littlewood},
 journal = {Physical Review Letters},
 volume = {122},
 number = {18},
 pages = {185301},
 year = {2019},
 month = may,
 publisher = {American Physical Society ({APS})}
}

@article{Hanai2020,
 doi = {10.1103/physrevresearch.2.033018},
 title = {Critical fluctuations at a many-body exceptional point},
 author = {Ryo Hanai and Peter B. Littlewood},
 journal = {Physical Review Research},
 volume = {2},
 number = {3},
 pages = {033018},
 year = {2020},
 month = jul,
 publisher = {American Physical Society ({APS})}
}

@article{ottLongTimeEvolution2009,
  title = {Long Time Evolution of Phase Oscillator Systems},
  author = {Ott, Edward and Antonsen, Thomas M.},
  year = 2009,
  month = may,
  journal = {Chaos: An Interdisciplinary Journal of Nonlinear Science},
  volume = {19},
  number = {2},
  pages = {023117},
  issn = {1054-1500},
  doi = {10.1063/1.3136851},
  urldate = {2025-10-21}
}

@article{ottLowDimensional2008a,
  title = {Low Dimensional Behavior of Large Systems of Globally Coupled Oscillators},
  author = {Ott, Edward and Antonsen, Thomas M.},
  year = 2008,
  month = sep,
  journal = {Chaos: An Interdisciplinary Journal of Nonlinear Science},
  volume = {18},
  number = {3},
  pages = {037113},
  issn = {1054-1500},
  doi = {10.1063/1.2930766},
  urldate = {2025-10-21}
}

@book{Ott2002,
  title={Chaos in Dynamical Systems},
  author={Ott, E.},
  isbn={9781139936576},
  year={2002},
  publisher={Cambridge University Press}
}

@book{Kuramoto1984,
     title = {Chemical Oscillations, Waves, And Turbulence},
    author = {Yoshiki Kuramoto},
      isbn = {9780486428819},
      year = {1984},
 publisher = {Springer},
 doi={10.1007/978-3-642-69689-3},
}

@misc{NIST_DLMF,
         key = "{\relax DLMF}",
       title = "{\it NIST Digital Library of Mathematical Functions}",
year={2023},
howpublished = "\url{https://dlmf.nist.gov/}, Release 1.1.10 of 2023-06-15",
         url = "https://dlmf.nist.gov/",
        note = "F.~W.~J. Olver, A.~B. {Olde Daalhuis}, D.~W. Lozier, B.~I. Schneider,
                R.~F. Boisvert, C.~W. Clark, B.~R. Miller, B.~V. Saunders,
                H.~S. Cohl, and M.~A. McClain, eds."}

@article{Fruchart2023,
 doi = {10.1146/annurev-conmatphys-040821-125506},
 title = {Odd Viscosity and Odd Elasticity},
 author = {Fruchart, Michel and Scheibner, Colin and Vitelli, Vincenzo},
 journal = {Annual Review of Condensed Matter Physics},
 volume = {14},
 number = {1},
 pages = {471–510},
 year = {2023},
 month = {Mar},
 issn = {1947-5462},
 publisher = {Annual Reviews}
}

@article{balleriniInteractionRulingAnimal2008,
  title = {Interaction Ruling Animal Collective Behavior Depends on Topological Rather than Metric Distance: {{Evidence}} from a Field Study},
  shorttitle = {Interaction Ruling Animal Collective Behavior Depends on Topological Rather than Metric Distance},
  author = {Ballerini, M. and Cabibbo, N. and Candelier, R. and Cavagna, A. and Cisbani, E. and Giardina, I. and Lecomte, V. and Orlandi, A. and Parisi, G. and Procaccini, A. and Viale, M. and Zdravkovic, V.},
  year = {2008},
  month = jan,
  journal = {Proceedings of the National Academy of Sciences},
  volume = {105},
  number = {4},
  pages = {1232--1237},
  publisher = {Proceedings of the National Academy of Sciences},
  doi = {10.1073/pnas.0711437105},
  urldate = {2025-09-12}
}

@article{cestnikHierarchyExactLowDimensional2022,
  title = {Hierarchy of {{Exact Low-Dimensional Reductions}} for {{Populations}} of {{Coupled Oscillators}}},
  author = {Cestnik, Rok and Pikovsky, Arkady},
  year = {2022},
  month = feb,
  journal = {Physical Review Letters},
  volume = {128},
  number = {5},
  pages = {054101},
  publisher = {American Physical Society},
  doi = {10.1103/PhysRevLett.128.054101},
  urldate = {2025-09-12}
}

@article{kumarEfficientFlockingMetric2021,
  title = {Efficient Flocking: Metric versus Topological Interactions},
  shorttitle = {Efficient Flocking},
  author = {Kumar, Vijay and De, Rumi},
  year = {2021},
  month = sep,
  journal = {Royal Society Open Science},
  volume = {8},
  number = {9},
  pages = {202158},
  publisher = {Royal Society},
  doi = {10.1098/rsos.202158},
  urldate = {2025-09-12}
}

@book{gardinerStochasticMethodsHandbook2009,
  title = {Stochastic Methods: A Handbook for the Natural and Social Sciences},
  shorttitle = {Stochastic Methods},
  author = {Gardiner, C. W. and Gardiner, C. W.},
  year = {2009},
  series = {Springer Series in Synergetics},
  edition = {4th ed},
  publisher = {Springer},
  address = {Berlin},
  isbn = {978-3-540-70712-7},
  lccn = {QA274 .G37 2009}
}

@article{tyulkinaDynamicsNoisyOscillator2018,
  title = {Dynamics of {{Noisy Oscillator Populations}} beyond the {{Ott-Antonsen Ansatz}}},
  author = {Tyulkina, Irina V. and Goldobin, Denis S. and Klimenko, Lyudmila S. and Pikovsky, Arkady},
  year = {2018},
  month = jun,
  journal = {Physical Review Letters},
  volume = {120},
  number = {26},
  pages = {264101},
  publisher = {American Physical Society},
  doi = {10.1103/PhysRevLett.120.264101},
  urldate = {2025-09-24}
}

@article{allesinaStabilityCriteriaComplex2012,
  title = {Stability Criteria for Complex Ecosystems},
  author = {Allesina, Stefano and Tang, Si},
  year = {2012},
  month = mar,
  journal = {Nature},
  volume = {483},
  number = {7388},
  pages = {205--208},
  publisher = {Nature Publishing Group},
  issn = {1476-4687},
  doi = {10.1038/nature10832},
  urldate = {2024-12-19},
  copyright = {2012 Springer Nature Limited},
  langid = {english}
}

@article{brandenbourgerNonreciprocalRoboticMetamaterials2019,
  title = {Non-Reciprocal Robotic Metamaterials},
  author = {Brandenbourger, Martin and Locsin, Xander and Lerner, Edan and Coulais, Corentin},
  year = {2019},
  month = oct,
  journal = {Nature Communications},
  volume = {10},
  number = {1},
  pages = {4608},
  publisher = {Nature Publishing Group},
  issn = {2041-1723},
  doi = {10.1038/s41467-019-12599-3},
  urldate = {2024-12-19},
  copyright = {2019 The Author(s)},
  langid = {english}
}

@article{chiacchioNonreciprocalDickeModel2023,
  title = {Nonreciprocal {{Dicke Model}}},
  author = {Chiacchio, Ezequiel I. Rodr{\'i}guez and Nunnenkamp, Andreas and Brunelli, Matteo},
  year = {2023},
  month = sep,
  journal = {Physical Review Letters},
  volume = {131},
  number = {11},
  pages = {113602},
  publisher = {American Physical Society},
  doi = {10.1103/PhysRevLett.131.113602},
  urldate = {2024-12-19}
}

@article{demirDynamicsPatternFormation2020,
  title = {Dynamics of Pattern Formation and Emergence of Swarming in {{Caenorhabditis}} Elegans},
  author = {Demir, Esin and Yaman, Y Ilker and Basaran, Mustafa and Kocabas, Askin},
  editor = {Krishna, Sandeep and Walczak, Aleksandra M and Guttal, Vishwesha and Marenduzzo, Davide},
  year = {2020},
  month = apr,
  journal = {eLife},
  volume = {9},
  pages = {e52781},
  publisher = {eLife Sciences Publications, Ltd},
  issn = {2050-084X},
  doi = {10.7554/eLife.52781},
  urldate = {2024-12-19}
}

@misc{huangActivePatternFormation2024,
  title = {Active Pattern Formation Emergent from Single-Species Nonreciprocity},
  author = {Huang, Zhi-Feng and te Vrugt, Michael and Wittkowski, Raphael and L{\"o}wen, Hartmut},
  year = {2024},
  month = apr,
  number = {arXiv:2404.10093},
  eprint = {2404.10093},
  primaryclass = {cond-mat},
  publisher = {arXiv},
  doi = {10.48550/arXiv.2404.10093},
  urldate = {2024-12-20},
  archiveprefix = {arXiv}
}

@article{johnTravellingLipidDomains2005,
  title = {Travelling Lipid Domains in a Dynamic Model for Protein-Induced Pattern Formation in Biomembranes},
  author = {John, Karin and B{\"a}r, Markus},
  year = {2005},
  month = jun,
  journal = {Physical Biology},
  volume = {2},
  number = {2},
  pages = {123},
  issn = {1478-3975},
  doi = {10.1088/1478-3975/2/2/005},
  urldate = {2024-12-19},
  langid = {english}
}

@article{soniOddFreeSurface2019,
  title = {The Odd Free Surface Flows of a Colloidal Chiral Fluid},
  author = {Soni, Vishal and Bililign, Ephraim S. and Magkiriadou, Sofia and Sacanna, Stefano and Bartolo, Denis and Shelley, Michael J. and Irvine, William T. M.},
  year = {2019},
  month = nov,
  journal = {Nature Physics},
  volume = {15},
  number = {11},
  pages = {1188--1194},
  publisher = {Nature Publishing Group},
  issn = {1745-2481},
  doi = {10.1038/s41567-019-0603-8},
  urldate = {2024-12-19},
  copyright = {2019 The Author(s), under exclusive licence to Springer Nature Limited},
  langid = {english}
}

@article{tanOddDynamicsLiving2022,
  title = {Odd Dynamics of Living Chiral Crystals},
  author = {Tan, Tzer Han and Mietke, Alexander and Li, Junang and Chen, Yuchao and Higinbotham, Hugh and Foster, Peter J. and Gokhale, Shreyas and Dunkel, J{\"o}rn and Fakhri, Nikta},
  year = {2022},
  month = jul,
  journal = {Nature},
  volume = {607},
  number = {7918},
  pages = {287--293},
  issn = {1476-4687},
  doi = {10.1038/s41586-022-04889-6},
  langid = {english},
  pmid = {35831595}
}

@article{watanabeAbsenceQuantumTime2015,
  title = {Absence of {{Quantum Time Crystals}}},
  author = {Watanabe, Haruki and Oshikawa, Masaki},
  year = {2015},
  month = jun,
  journal = {Physical Review Letters},
  volume = {114},
  number = {25},
  pages = {251603},
  publisher = {American Physical Society},
  doi = {10.1103/PhysRevLett.114.251603},
  urldate = {2024-12-19}
}

@article{zelleUniversalPhenomenologyCritical2024,
  title = {Universal {{Phenomenology}} at {{Critical Exceptional Points}} of {{Nonequilibrium O}} ( {{N}} ) {{Models}}},
  author = {Zelle, Carl Philipp and Daviet, Romain and Rosch, Achim and Diehl, Sebastian},
  year = {2024},
  month = jun,
  journal = {Physical Review X},
  volume = {14},
  number = {2},
  pages = {021052},
  issn = {2160-3308},
  doi = {10.1103/PhysRevX.14.021052},
  urldate = {2024-12-19},
  langid = {english}
}

@article{hanaiNonreciprocalFrustrationTime2024,
  title = {Nonreciprocal {{Frustration}}: {{Time Crystalline Order-by-Disorder Phenomenon}} and a {{Spin-Glass-like State}}},
  shorttitle = {Nonreciprocal {{Frustration}}},
  author = {Hanai, Ryo},
  year = {2024},
  month = feb,
  journal = {Physical Review X},
  volume = {14},
  number = {1},
  pages = {011029},
  issn = {2160-3308},
  doi = {10.1103/PhysRevX.14.011029},
  urldate = {2024-04-25},
  langid = {english}
}

@misc{avniNonreciprocalIsingModel2024b,
  title = {The Non-Reciprocal {{Ising}} Model},
  author = {Avni, Yael and Fruchart, Michel and Martin, David and Seara, Daniel and Vitelli, Vincenzo},
  year = {2024},
  month = sep,
  number = {arXiv:2311.05471},
  eprint = {2311.05471},
  primaryclass = {cond-mat},
  publisher = {arXiv},
  doi = {10.48550/arXiv.2311.05471},
  urldate = {2024-12-19},
  archiveprefix = {arXiv}
}

@article{belyanskyPhaseTransitionsNonreciprocal2025,
  title = {Phase {{Transitions}} in {{Nonreciprocal Driven-Dissipative Condensates}}},
  author = {Belyansky, Ron and Weis, Cheyne and Hanai, Ryo and Littlewood, Peter B. and Clerk, Aashish A.},
  year = 2025,
  month = sep,
  journal = {Physical Review Letters},
  volume = {135},
  number = {12},
  pages = {123401},
  publisher = {American Physical Society},
  doi = {10.1103/gphr-d1bc},
  urldate = {2025-11-18}
}

@article{hanaiPhotoinducedNonreciprocalMagnetism2024,
  title = {Photoinduced Non-Reciprocal Magnetism},
  author = {Hanai, Ryo and Ootsuki, Daiki and Tazai, Rina},
  year = 2025,
  month = sep,
  journal = {Nature Communications},
  volume = {16},
  number = {1},
  pages = {8195},
  publisher = {Nature Publishing Group},
  issn = {2041-1723},
  doi = {10.1038/s41467-025-62707-9},
  urldate = {2025-11-18},
  copyright = {2025 The Author(s)},
  langid = {english}
}

@article{nadolnyNonreciprocalSynchronizationActive2024,
  title = {Nonreciprocal {{Synchronization}} of {{Active Quantum Spins}}},
  author = {Nadolny, Tobias and Bruder, Christoph and Brunelli, Matteo},
  year = 2025,
  month = jan,
  journal = {Physical Review X},
  volume = {15},
  number = {1},
  pages = {011010},
  publisher = {American Physical Society},
  doi = {10.1103/PhysRevX.15.011010},
  urldate = {2025-11-18}
}

@article{parkavousiEnhancedStabilityChaotic2024,
  title = {Enhanced {{Stability}} and {{Chaotic Condensates}} in {{Multispecies Nonreciprocal Mixtures}}},
  author = {Parkavousi, Laya and Rana, Navdeep and Golestanian, Ramin and Saha, Suropriya},
  year = 2025,
  month = apr,
  journal = {Physical Review Letters},
  volume = {134},
  number = {14},
  pages = {148301},
  publisher = {American Physical Society},
  doi = {10.1103/PhysRevLett.134.148301},
  urldate = {2025-11-18}
}

@article{weisExceptionalPointsNonlinear2023a,
  title = {Generalized Exceptional Points in Nonlinear and Stochastic Dynamics},
  author = {Weis, Cheyne and Fruchart, Michel and Hanai, Ryo and Kawagoe, Kyle and Littlewood, Peter B. and Vitelli, Vincenzo},
  year = 2025,
  month = nov,
  journal = {Physical Review Research},
  volume = {7},
  number = {4},
  pages = {043157},
  publisher = {American Physical Society},
  doi = {10.1103/mnn4-b298},
  urldate = {2025-11-18}
}

@article{braunsNonreciprocalPatternFormation2024,
  title = {Nonreciprocal {{Pattern Formation}} of {{Conserved Fields}}},
  author = {Brauns, Fridtjof and Marchetti, M. Cristina},
  year = {2024},
  month = apr,
  journal = {Phys. Rev. X},
  volume = {14},
  number = {2},
  pages = {021014},
  issn = {2160-3308},
  doi = {10.1103/PhysRevX.14.021014},
  urldate = {2024-12-19},
  langid = {english}
}

@article{Dadhichi2020,
  title = {Nonmutual torques and the unimportance of motility for long-range order in two-dimensional flocks},
  author = {Dadhichi, Lokrshi Prawar and Kethapelli, Jitendra and Chajwa, Rahul and Ramaswamy, Sriram and Maitra, Ananyo},
  journal = {Phys. Rev. E},
  volume = {101},
  issue = {5},
  pages = {052601},
  numpages = {13},
  year = {2020},
  month = {May},
  publisher = {American Physical Society},
  doi = {10.1103/PhysRevE.101.052601},
  url = {https://link.aps.org/doi/10.1103/PhysRevE.101.052601}
}

@article{Loos2023,
  title = {Long-Range Order and Directional Defect Propagation in the Nonreciprocal $\mathit{XY}$ Model with Vision Cone Interactions},
  author = {Loos, Sarah A. M. and Klapp, Sabine H. L. and Martynec, Thomas},
  journal = {Phys. Rev. Lett.},
  volume = {130},
  issue = {19},
  pages = {198301},
  numpages = {6},
  year = {2023},
  month = {May},
  publisher = {American Physical Society},
  doi = {10.1103/PhysRevLett.130.198301},
  url = {https://link.aps.org/doi/10.1103/PhysRevLett.130.198301}
}

@article{markovichNonreciprocityOddViscosity2024,
  title = {Nonreciprocity and Odd Viscosity in Chiral Active Fluids},
  author = {Markovich, Tomer and Lubensky, Tom C.},
  year = {2024},
  month = may,
  journal = {Proc. Natl. Acad. Sci.},
  volume = {121},
  number = {19},
  pages = {e2219385121},
  publisher = {Proc. Natl. Acad. Sci.},
  doi = {10.1073/pnas.2219385121},
  urldate = {2025-01-13}
}

@article{muruganTopologicallyProtectedModes2017,
  title = {Topologically Protected Modes in Non-Equilibrium Stochastic Systems},
  author = {Murugan, Arvind and Vaikuntanathan, Suriyanarayanan},
  year = {2017},
  month = jan,
  journal = {Nat Commun.},
  volume = {8},
  number = {1},
  pages = {13881},
  publisher = {Nature Publishing Group},
  issn = {2041-1723},
  doi = {10.1038/ncomms13881},
  urldate = {2025-01-13},
  copyright = {2017 The Author(s)},
  langid = {english}
}

@article{Pisegna2024,
  title = {Emergent Polar Order in Nonpolar Mixtures with Nonreciprocal Interactions},
  year = {2024},
  month = dec,
  journal = {Proc. Natl. Acad. Sci.},
  volume = {121},
  number = {51},
  pages = {e2407705121},
  publisher = {Proceedings of the National Academy of Sciences},
  doi = {10.1073/pnas.2407705121},
  urldate = {2025-02-04},
  author = {Pisegna, Giulia and Saha, Suropriya and Golestanian, Ramin}
}

@article{sompolinskyTemporalAssociationAsymmetric1986,
  title = {Temporal {{Association}} in {{Asymmetric Neural Networks}}},
  author = {Sompolinsky, H. and Kanter, I.},
  year = {1986},
  month = dec,
  journal = {Phys. Rev. Lett},
  volume = {57},
  number = {22},
  pages = {2861--2864},
  publisher = {American Physical Society},
  doi = {10.1103/PhysRevLett.57.2861},
  urldate = {2025-01-11}
}

@article{veenstraNonreciprocalTopologicalSolitons2024,
  title = {Non-Reciprocal Topological Solitons in Active Metamaterials},
  author = {Veenstra, Jonas and Gamayun, Oleksandr and Guo, Xiaofei and Sarvi, Anahita and Meinersen, Chris Ventura and Coulais, Corentin},
  year = {2024},
  month = mar,
  journal = {Nature},
  volume = {627},
  number = {8004},
  pages = {528--533},
  publisher = {Nature Publishing Group},
  issn = {1476-4687},
  doi = {10.1038/s41586-024-07097-6},
  urldate = {2025-01-13},
  copyright = {2024 The Author(s), under exclusive licence to Springer Nature Limited},
  langid = {english}
}

@article{youNonreciprocityGenericRoute2020,
  title = {Nonreciprocity as a Generic Route to Traveling States},
  author = {You, Zhihong and Baskaran, Aparna and Marchetti, M. Cristina},
  year = {2020},
  month = aug,
  journal = {Proc. Natl. Acad. Sci.},
  volume = {117},
  number = {33},
  pages = {19767--19772},
  publisher = {Proceedings of the National Academy of Sciences},
  doi = {10.1073/pnas.2010318117},
  urldate = {2025-01-11}
}

@article{zhengTopologicalMechanismRobust2024,
  title = {A Topological Mechanism for Robust and Efficient Global Oscillations in Biological Networks},
  author = {Zheng, Chongbin and Tang, Evelyn},
  year = {2024},
  month = jul,
  journal = {Nat Commun},
  volume = {15},
  number = {1},
  pages = {6453},
  publisher = {Nature Publishing Group},
  issn = {2041-1723},
  doi = {10.1038/s41467-024-50510-x},
  urldate = {2025-01-13},
  copyright = {2024 The Author(s)},
  langid = {english}
}

@article{bowickSymmetryThermodynamicsTopology2022,
  title = {Symmetry, {{Thermodynamics}}, and {{Topology}} in {{Active Matter}}},
  author = {Bowick, Mark J. and Fakhri, Nikta and Marchetti, M. Cristina and Ramaswamy, Sriram},
  year = {2022},
  month = feb,
  journal = {Phys. Rev. X},
  volume = {12},
  number = {1},
  pages = {010501},
  issn = {2160-3308},
  doi = {10.1103/PhysRevX.12.010501},
  urldate = {2025-01-22},
  langid = {english}
}

@misc{khassehActiveQuantumFlocks2024,
  title = {Active Quantum Flocks},
  author = {Khasseh, Reyhaneh and Wald, Sascha and Moessner, Roderich and Weber, Christoph A. and Heyl, Markus},
  year = {2024},
  month = sep,
  number = {arXiv:2308.01603},
  eprint = {2308.01603},
  primaryclass = {quant-ph},
  publisher = {arXiv},
  doi = {10.48550/arXiv.2308.01603},
  urldate = {2025-02-16},
  archiveprefix = {arXiv}
}

@article{weiderpassSolvingKineticIsing2025,
  title = {Solving the Kinetic {{Ising}} Model with Nonreciprocity},
  author = {Weiderpass, Gabriel Artur and Sharma, Mayur and Sethi, Savdeep},
  year = {2025},
  month = feb,
  journal = {Physical Review E},
  volume = {111},
  number = {2},
  pages = {024107},
  publisher = {American Physical Society},
  doi = {10.1103/PhysRevE.111.024107},
  urldate = {2025-02-16}
}

@article{godrecheDynamicsDirectedIsing2011,
  title = {Dynamics of the Directed {{Ising}} Chain},
  author = {Godr{\`e}che, Claude},
  year = {2011},
  month = apr,
  journal = {Journal of Statistical Mechanics: Theory and Experiment},
  volume = {2011},
  number = {04},
  pages = {P04005},
  issn = {1742-5468},
  doi = {10.1088/1742-5468/2011/04/P04005},
  urldate = {2025-02-16},
  langid = {english}
}

@article{searaNonreciprocalInteractionsSpatially2023,
  title = {Non-Reciprocal Interactions Spatially Propagate Fluctuations in a {{2D Ising}} Model},
  author = {Seara, Daniel S and Piya, Akash and Tabatabai, A Pasha},
  year = {2023},
  month = apr,
  journal = {Journal of Statistical Mechanics: Theory and Experiment},
  volume = {2023},
  number = {4},
  pages = {043209},
  publisher = {IOP Publishing},
  issn = {1742-5468},
  doi = {10.1088/1742-5468/accce7},
  urldate = {2025-02-16},
  langid = {english}
}

@misc{bhattEmergentHydrodynamicsNonreciprocal2023,
  title = {Emergent Hydrodynamics in a Non-Reciprocal Classical Isotropic Magnet},
  author = {Bhatt, Nisarg and Mukerjee, Subroto and Ramaswamy, Sriram},
  year = {2023},
  month = dec,
  number = {arXiv:2312.16500},
  eprint = {2312.16500},
  primaryclass = {cond-mat},
  publisher = {arXiv},
  doi = {10.48550/arXiv.2312.16500},
  urldate = {2025-02-22},
  archiveprefix = {arXiv}
}

@article{dasDrivenHeisenbergMagnets2002,
  title = {Driven {{Heisenberg}} Magnets: {{Nonequilibrium}} Criticality, Spatiotemporal Chaos and Control},
  shorttitle = {Driven {{Heisenberg}} Magnets},
  author = {Das, J. and Rao, M. and Ramaswamy, S.},
  year = {2002},
  month = nov,
  journal = {Europhysics Letters},
  volume = {60},
  number = {3},
  pages = {418},
  publisher = {IOP Publishing},
  issn = {0295-5075},
  doi = {10.1209/epl/i2002-00280-2},
  urldate = {2025-02-22},
  langid = {english}
}

@article{Acebron2005,
  title = {The {{Kuramoto}} Model: {{A}} Simple Paradigm for Synchronization Phenomena},
  shorttitle = {The {{Kuramoto}} Model},
  author = {Acebr{\'o}n, Juan A. and Bonilla, L. L. and P{\'e}rez Vicente, Conrad J. and Ritort, F{\'e}lix and Spigler, Renato},
  year = {2005},
  month = apr,
  journal = {Reviews of Modern Physics},
  volume = {77},
  number = {1},
  pages = {137--185},
  issn = {0034-6861, 1539-0756},
  doi = {10.1103/RevModPhys.77.137},
  urldate = {2025-03-03},
  copyright = {http://link.aps.org/licenses/aps-default-license},
  langid = {english}
}

@article{blumenthalPhaseTransitionChaos2024,
  title = {Phase {{Transition}} to {{Chaos}} in {{Complex Ecosystems}} with {{Nonreciprocal Species-Resource Interactions}}},
  author = {Blumenthal, Emmy and Rocks, Jason W. and Mehta, Pankaj},
  year = {2024},
  month = mar,
  journal = {Physical Review Letters},
  volume = {132},
  number = {12},
  pages = {127401},
  publisher = {American Physical Society},
  doi = {10.1103/PhysRevLett.132.127401},
  urldate = {2025-04-07}
}

@article{faugerasConstructiveMeanfieldAnalysis2009,
  title = {A Constructive Mean-Field Analysis of Multi Population Neural Networks with Random Synaptic Weights and Stochastic Inputs},
  author = {Faugeras, Olivier D. and Touboul, Jonathan D. and Cessac, Bruno},
  year = {2009},
  month = feb,
  journal = {Frontiers in Computational Neuroscience},
  volume = {3},
  publisher = {Frontiers},
  issn = {1662-5188},
  doi = {10.3389/neuro.10.001.2009},
  urldate = {2025-04-07},
  langid = {english}
}

@article{grafThermodynamicStabilityCritical2022,
  title = {Thermodynamic Stability and Critical Points in Multicomponent Mixtures with Structured Interactions},
  author = {Graf, Isabella R. and Machta, Benjamin B.},
  year = {2022},
  month = aug,
  journal = {Physical Review Research},
  volume = {4},
  number = {3},
  pages = {033144},
  publisher = {American Physical Society},
  doi = {10.1103/PhysRevResearch.4.033144},
  urldate = {2025-04-07}
}

@article{huEmergentPhasesEcological2022,
  title = {Emergent Phases of Ecological Diversity and Dynamics Mapped in Microcosms},
  author = {Hu, Jiliang and Amor, Daniel R. and Barbier, Matthieu and Bunin, Guy and Gore, Jeff},
  year = {2022},
  month = oct,
  journal = {Science},
  volume = {378},
  number = {6615},
  pages = {85--89},
  publisher = {American Association for the Advancement of Science},
  doi = {10.1126/science.abm7841},
  urldate = {2025-04-07}
}

@article{jacobsPhaseTransitionsBiological2017,
  title = {Phase {{Transitions}} in {{Biological Systems}} with {{Many Components}}},
  author = {Jacobs, William M. and Frenkel, Daan},
  year = {2017},
  month = feb,
  journal = {Biophysical Journal},
  volume = {112},
  number = {4},
  pages = {683--691},
  issn = {00063495},
  doi = {10.1016/j.bpj.2016.10.043},
  urldate = {2025-04-07},
  langid = {english}
}

@article{jacobsSelfAssemblyBiomolecularCondensates2021,
  title = {Self-{{Assembly}} of {{Biomolecular Condensates}} with {{Shared Components}}},
  author = {Jacobs, William M.},
  year = {2021},
  month = jun,
  journal = {Physical Review Letters},
  volume = {126},
  number = {25},
  pages = {258101},
  publisher = {American Physical Society},
  doi = {10.1103/PhysRevLett.126.258101},
  urldate = {2025-04-07}
}

@article{kimDynamicsMultipleInteracting2020,
  title = {Dynamics of Multiple Interacting Excitatory and Inhibitory Populations with Delays},
  author = {Kim, Christopher M. and Egert, Ulrich and Kumar, Arvind},
  year = {2020},
  month = aug,
  journal = {Physical Review E},
  volume = {102},
  number = {2},
  pages = {022308},
  publisher = {American Physical Society},
  doi = {10.1103/PhysRevE.102.022308},
  urldate = {2025-04-06}
}

@misc{palmigianoCommonRulesUnderlying2023,
  title = {Common Rules Underlying Optogenetic and Behavioral Modulation of Responses in Multi-Cell-Type {{V1}} Circuits},
  author = {Palmigiano, Agostina and Fumarola, Francesco and Mossing, Daniel P. and Kraynyukova, Nataliya and Adesnik, Hillel and Miller, Kenneth D.},
  year = {2023},
  month = jan,
  primaryclass = {New Results},
  pages = {2020.11.11.378729},
  publisher = {bioRxiv},
  doi = {10.1101/2020.11.11.378729},
  urldate = {2025-04-07},
  archiveprefix = {bioRxiv},
  chapter = {New Results},
  copyright = {{\copyright} 2023, Posted by Cold Spring Harbor Laboratory. The copyright holder for this pre-print is the author. All rights reserved. The material may not be redistributed, re-used or adapted without the author's permission.},
  langid = {english}
}

@article{reichenbachMobilityPromotesJeopardizes2007,
  title = {Mobility Promotes and Jeopardizes Biodiversity in Rock--Paper--Scissors Games},
  author = {Reichenbach, Tobias and Mobilia, Mauro and Frey, Erwin},
  year = {2007},
  month = aug,
  journal = {Nature},
  volume = {448},
  number = {7157},
  pages = {1046--1049},
  publisher = {Nature Publishing Group},
  issn = {1476-4687},
  doi = {10.1038/nature06095},
  urldate = {2025-04-07},
  copyright = {2007 Springer Nature Limited},
  langid = {english}
}

@article{chenEmergenceMultiphaseCondensates2024,
  title = {Emergence of {{Multiphase Condensates}} from a {{Limited Set}} of {{Chemical Building Blocks}}},
  author = {Chen, Fan and Jacobs, William M.},
  year = {2024},
  month = aug,
  journal = {Journal of Chemical Theory and Computation},
  volume = {20},
  number = {15},
  pages = {6881--6889},
  publisher = {American Chemical Society},
  issn = {1549-9618},
  doi = {10.1021/acs.jctc.4c00323},
  urldate = {2025-04-08}
}

@article{xieSocialConsensusInfluence2011,
  title = {Social Consensus through the Influence of Committed Minorities},
  author = {Xie, J. and Sreenivasan, S. and Korniss, G. and Zhang, W. and Lim, C. and Szymanski, B. K.},
  year = {2011},
  month = jul,
  journal = {Physical Review E},
  volume = {84},
  number = {1},
  pages = {011130},
  publisher = {American Physical Society},
  doi = {10.1103/PhysRevE.84.011130},
  urldate = {2025-04-08}
}

@article{scheibnerOddElasticity2020,
  title = {Odd Elasticity},
  author = {Scheibner, Colin and Souslov, Anton and Banerjee, Debarghya and Sur{\'o}wka, Piotr and Irvine, William T. M. and Vitelli, Vincenzo},
  year = {2020},
  month = apr,
  journal = {Nature Physics},
  volume = {16},
  number = {4},
  pages = {475--480},
  publisher = {Nature Publishing Group},
  issn = {1745-2481},
  doi = {10.1038/s41567-020-0795-y},
  urldate = {2025-04-25},
  copyright = {2020 The Author(s), under exclusive licence to Springer Nature Limited},
  langid = {english}
}

@article{lotkaAnalyticalNoteCertain1920,
  title = {Analytical {{Note}} on {{Certain Rhythmic Relations}} in {{Organic Systems}}},
  author = {Lotka, Alfred J.},
  year = {1920},
  month = jul,
  journal = {Proceedings of the National Academy of Sciences},
  volume = {6},
  number = {7},
  pages = {410--415},
  publisher = {Proceedings of the National Academy of Sciences},
  doi = {10.1073/pnas.6.7.410},
  urldate = {2025-05-06}
}

@article{nowakEvolutionaryGamesSpatial1992,
  title = {Evolutionary Games and Spatial Chaos},
  author = {Nowak, Martin A. and May, Robert M.},
  year = {1992},
  month = oct,
  journal = {Nature},
  volume = {359},
  number = {6398},
  pages = {826--829},
  publisher = {Nature Publishing Group},
  issn = {1476-4687},
  doi = {10.1038/359826a0},
  urldate = {2025-05-06},
  copyright = {1992 Springer Nature Limited},
  langid = {english}
}

@article{rosenzweigGraphicalRepresentationStability1963,
  title = {Graphical {{Representation}} and {{Stability Conditions}} of {{Predator-Prey Interactions}}},
  author = {Rosenzweig, M. L. and MacArthur, R. H.},
  year = {1963},
  month = jul,
  journal = {The American Naturalist},
  volume = {97},
  number = {895},
  pages = {209--223},
  publisher = {The University of Chicago Press},
  issn = {0003-0147},
  doi = {10.1086/282272},
  urldate = {2025-05-06}
}

@article{wilsonExcitatoryInhibitoryInteractions1972,
  title = {Excitatory and {{Inhibitory Interactions}} in {{Localized Populations}} of {{Model Neurons}}},
  author = {Wilson, Hugh R. and Cowan, Jack D.},
  year = {1972},
  month = jan,
  journal = {Biophysical Journal},
  volume = {12},
  number = {1},
  pages = {1--24},
  issn = {0006-3495},
  doi = {10.1016/S0006-3495(72)86068-5},
  urldate = {2025-05-06}
}

@misc{khemaniBriefHistoryTime2019,
  title = {A {{Brief History}} of {{Time Crystals}}},
  author = {Khemani, Vedika and Moessner, Roderich and Sondhi, S. L.},
  year = {2019},
  month = oct,
  number = {arXiv:1910.10745},
  eprint = {1910.10745},
  primaryclass = {cond-mat},
  publisher = {arXiv},
  doi = {10.48550/arXiv.1910.10745},
  urldate = {2025-05-07},
  archiveprefix = {arXiv}
}

@article{kongkhambutObservationContinuousTime2022,
  title = {Observation of a Continuous Time Crystal},
  author = {Kongkhambut, Phatthamon and Skulte, Jim and Mathey, Ludwig and Cosme, Jayson G. and Hemmerich, Andreas and Ke{\ss}ler, Hans},
  year = {2022},
  month = aug,
  journal = {Science},
  volume = {377},
  number = {6606},
  pages = {670--673},
  publisher = {American Association for the Advancement of Science},
  doi = {10.1126/science.abo3382},
  urldate = {2025-05-07}
}

@book{hatcherAlgebraicTopology2001,
  title = {Algebraic Topology},
  author = {Hatcher, Allen},
  year = {2001},
  publisher = {Cambridge university press},
  address = {New York},
  isbn = {978-0-521-79540-1},
  langid = {english},
  lccn = {514.2}
}

@article{Shilnikov1965,
  author       = {Shilnikov, Leonid Pavlovich},
  title        = {A Case of the Existence of a Denumerable Set of Periodic Motions},
  journal      = {Doklady Akademii Nauk SSSR},
  year         = {1965},
  volume       = {160},
  number       = {3},
  pages        = {558--561},
  language     = {Russian},
  note         = {English translation: \emph{Soviet Math. Dokl.} \textbf{6} (1965), 163--166},
  url          = {http://mi.mathnet.ru/eng/dan30608},
  mrnumber     = {0173047},
  zbnumber     = {0136.08202}
}

@article{Shilnikov1967,
  author       = {Shilnikov, Leonid Pavlovich},
  title        = {Existence of a Countable Set of Periodic Motions in a Four-Dimensional Space in an Extended Neighborhood of a Saddle–Focus},
  journal      = {Doklady Akademii Nauk SSSR},
  year         = {1967},
  volume       = {172},
  number       = {1},
  pages        = {54--57},
  language     = {Russian},
  note         = {English translation: \emph{Soviet Math. Dokl.} \textbf{8} (1967), 54--58},
  url          = {http://mi.mathnet.ru/eng/dan32794},
  mrnumber     = {0210987},
  zbnumber     = {0155.41805}
}

@article{fowlerBifocalHomoclinicOrbits1991,
  title = {Bifocal Homoclinic Orbits in Four Dimensions},
  author = {Fowler, A. C. and Sparrow, C. T.},
  year = {1991},
  month = nov,
  journal = {Nonlinearity},
  volume = {4},
  number = {4},
  pages = {1159},
  issn = {0951-7715},
  doi = {10.1088/0951-7715/4/4/007},
  urldate = {2025-06-23},
  langid = {english}
}

@article{gaspardWhatCanWe1983,
  title = {What Can We Learn from Homoclinic Orbits in Chaotic Dynamics?},
  author = {Gaspard, P. and Nicolis, G.},
  year = {1983},
  month = jun,
  journal = {Journal of Statistical Physics},
  volume = {31},
  number = {3},
  pages = {499--518},
  issn = {1572-9613},
  doi = {10.1007/BF01019496},
  urldate = {2025-06-23},
  langid = {english}
}

@book{homburgGlobalAspectsHomoclinic1996,
  title = {Global Aspects of Homoclinic Bifurcations of Vector Fields},
  author = {Homburg, Ale Jan},
  year = {1996},
  series = {Memoirs of the {{American Mathematical Society}}, {{Volume}} 121, {{Number}} 578},
  edition = {1st ed.},
  publisher = {American Mathematical Society},
  address = {Providence, Rhode Island},
  isbn = {978-1-4704-0163-4},
  langid = {english}
}

@article{sandstedeCenterManifoldsHomoclinic2000,
  title = {Center {{Manifolds}} for {{Homoclinic Solutions}}},
  author = {Sandstede, Bj{\"o}rn},
  year = {2000},
  month = jul,
  journal = {Journal of Dynamics and Differential Equations},
  volume = {12},
  number = {3},
  pages = {449--510},
  issn = {1572-9222},
  doi = {10.1023/A:1026412926537},
  urldate = {2025-06-23},
  langid = {english}
}

@book{shilnikovMethodsQualitativeTheory1998,
  author    = {Shilnikov, Leonid P. and Shilnikov, Andrey L. and Turaev, Dmitry V.},
  title     = {Methods of Qualitative Theory in Nonlinear Dynamics: {Part I}},
  year      = {1998},
  series    = {World Scientific Series on Nonlinear Science, Series A},
  volume    = {4},
  publisher = {World Scientific},
  address   = {Singapore; River Edge, NJ},
  isbn      = {9810233825},
  pages     = {416},
  doi       = {10.1142/3707}
}

@book{shilnikovMethodsQualitativeTheory2001a,
  author    = {Shilnikov, Leonid P. and Shilnikov, Andrey L. and Turaev, Dmitry V. and Chua, Leon O.},
  title     = {Methods of Qualitative Theory in Nonlinear Dynamics: {Part II}},
  year      = {2001},
  series    = {World Scientific Series on Nonlinear Science, Series A},
  volume    = {5},
  publisher = {World Scientific},
  address   = {Singapore; River Edge, NJ},
  pages     = {592},
  isbn      = {9810240724},
  isbn13    = {9789810240721},
  doi       = {10.1142/4221}
}

@phdthesis{turaevBifurcationsDynamicalSystems1991,
  title = {{On bifurcations of dynamical systems with two homoclinic curves of the saddle}},
  author = {Turaev, D.V.},
  year = {1991},
  langid = {russian},
  school = {Gorkii State University}
}

@article{xingOrderedIntricacyShilnikov2021,
  title = {Ordered Intricacy of {{Shilnikov}} Saddle-Focus Homoclinics in Symmetric Systems},
  author = {Xing, Tingli and Pusuluri, Krishna and Shilnikov, Andrey L.},
  year = {2021},
  month = jul,
  journal = {Chaos: An Interdisciplinary Journal of Nonlinear Science},
  volume = {31},
  number = {7},
  pages = {073143},
  issn = {1054-1500},
  doi = {10.1063/5.0054776},
  urldate = {2025-06-23}
}

@article{gironSynchronizationUnveilsOrganization2016a,
  title = {Synchronization Unveils the Organization of Ecological Networks with Positive and Negative Interactions},
  author = {Gir{\'o}n, Andrea and Saiz, Hugo and Bacelar, Flora S. and Andrade, Roberto F. S. and {G{\'o}mez-Garde{\~n}es}, Jes{\'u}s},
  year = {2016},
  month = jun,
  journal = {Chaos (Woodbury, N.Y.)},
  volume = {26},
  number = {6},
  pages = {065302},
  issn = {1089-7682},
  doi = {10.1063/1.4952960},
  langid = {english},
  pmid = {27368792}
}

@article{vandermeerNewFormsStructure,
  title = {New Forms of Structure in Ecosystems Revealed with the {{Kuramoto}} Model},
  author = {Vandermeer, John and {Hajian-Forooshani}, Zachary and Medina, Nicholas and Perfecto, Ivette},
  year={2021},
  journal = {Royal Society Open Science},
  volume = {8},
  number = {3},
  pages = {210122},
  issn = {2054-5703},
  doi = {10.1098/rsos.210122},
  urldate = {2025-06-24},
  pmcid = {PMC8074911},
  pmid = {33959373}
}

@book{bredonTopologyGeometry2005,
  title = {Topology and Geometry},
  author = {Bredon, Glen E.},
  year = {2005},
  series = {Graduate Texts in Mathematics},
  edition = {7. print.},
  number = {139},
  publisher = {Springer},
  address = {New York},
  isbn = {978-0-387-97926-7 978-3-540-97926-5 978-1-4419-3103-0},
  langid = {english}
}

@book{frankelGeometryPhysicsIntroduction2012,
  title = {The Geometry of Physics: An Introduction},
  shorttitle = {The Geometry of Physics},
  author = {Frankel, Theodore},
  year = {2012},
  edition = {Third edition},
  publisher = {Cambridge University Press},
  address = {Cambridge},
  doi = {10.1017/CBO9781139061377},
  isbn = {978-1-139-06137-7 978-1-107-01673-6 978-1-107-60260-1},
  langid = {english}
}

@book{arnoldAbelianGroupsRepresentations2000,
  title = {Abelian Groups and Representations of Finite Partially Ordered Sets},
  author = {Arnold, David M.},
  year = {2000},
  series = {{{CMS}} Books in Mathematics},
  number = {2},
  publisher = {Springer},
  address = {New York},
  isbn = {978-1-4419-8750-1},
  langid = {english}
}

@book{birmanBraidsLinksMapping2016,
  title = {Braids, {{Links}}, and {{Mapping Class Groups}}: {{AM-82}}},
  shorttitle = {Braids, {{Links}}, and {{Mapping Class Groups}}},
  author = {Birman, Joan S.},
  year = {2016},
  series = {Annals of {{Mathematics Studies}}},
  number = {82},
  publisher = {Princeton University Press},
  address = {Princeton, NJ},
  doi = {10.1515/9781400881420},
  isbn = {978-0-691-08149-6 978-1-4008-8142-0},
  langid = {english}
}

@book{farbPrimerMappingClass2012,
  title = {A Primer on Mapping Class Groups},
  author = {Farb, Benson and Margalit, Dan},
  year = {2012},
  series = {Princeton Mathematical Series},
  number = {49},
  publisher = {Princeton University Press},
  address = {Princeton Oxford},
  doi = {10.1515/9781400839049},
  isbn = {978-0-691-14794-9 978-1-4008-3904-9},
  langid = {english}
}

@book{schrijverTheoryLinearInteger2011,
  title = {Theory of Linear and Integer Programming},
  author = {Schrijver, Alexander},
  year = {2011},
  series = {Wiley-{{Interscience}} Series in Discrete Mathematics and Optimization},
  edition = {Nachdr.},
  publisher = {Wiley},
  address = {Chichester Weinheim},
  isbn = {978-0-471-98232-6},
  langid = {english}
}

@book{barenblattNonlinearDynamicsTurbulence1983,
  title     = {Nonlinear Dynamics and Turbulence},
  author= {Barenblatt, G. I. and Iooss, Gérard and Joseph, Daniel D.},
  year      = {1983},
  series    = {Interaction of Mechanics and Mathematics Series},
  publisher = {Pitman Advanced Publishing Program},
  address   = {Boston},
  pages     = {356},
  isbn      = {978-0273085607}
}

@article{ovsyannikovSystemsHomoclinicCurve1991,
  title = {Systems with a Homoclinic Curve of Multidimensional Saddle-focus, and Spiral Chaos},
  author = {Ovsyannikov, I. M. and Shil{$\prime$}nikov, L.P.},
  year = {1991},
  journal = {Matematicheskii Sbornik},
  volume = {182},
  number = {7},
  pages = {1043--1073},
  doi = {10.1070/SM1992v073n02ABEH002553}
}

@article{ashwinMathematicalFrameworksOscillatory2016,
  title = {Mathematical {{Frameworks}} for {{Oscillatory Network Dynamics}} in {{Neuroscience}}},
  author = {Ashwin, Peter and Coombes, Stephen and Nicks, Rachel},
  year = {2016},
  month = jan,
  journal = {The Journal of Mathematical Neuroscience},
  volume = {6},
  number = {1},
  pages = {2},
  issn = {2190-8567},
  doi = {10.1186/s13408-015-0033-6},
  urldate = {2025-07-12}
}

@article{bickUnderstandingDynamicsBiological2020,
  title = {Understanding the Dynamics of Biological and Neural Oscillator Networks through Exact Mean-Field Reductions: A Review},
  shorttitle = {Understanding the Dynamics of Biological and Neural Oscillator Networks through Exact Mean-Field Reductions},
  author = {Bick, Christian and Goodfellow, Marc and Laing, Carlo R. and Martens, Erik A.},
  year = {2020},
  month = may,
  journal = {The Journal of Mathematical Neuroscience},
  volume = {10},
  number = {1},
  pages = {9},
  issn = {2190-8567},
  doi = {10.1186/s13408-020-00086-9},
  urldate = {2025-07-12}
}

@article{weerasinghePredictingEffectsDeep2019,
  title = {Predicting the Effects of Deep Brain Stimulation Using a Reduced Coupled Oscillator Model},
  author = {Weerasinghe, Gihan and Duchet, Benoit and Cagnan, Hayriye and Brown, Peter and Bick, Christian and Bogacz, Rafal},
  year = {2019},
  month = aug,
  journal = {PLOS Computational Biology},
  volume = {15},
  number = {8},
  pages = {e1006575},
  publisher = {Public Library of Science},
  issn = {1553-7358},
  doi = {10.1371/journal.pcbi.1006575},
  urldate = {2025-07-12},
  langid = {english}
}

@article{wiesenfeldFrequencyLockingJosephson1998,
  title = {Frequency Locking in {{Josephson}} Arrays: {{Connection}} with the {{Kuramoto}} Model},
  shorttitle = {Frequency Locking in {{Josephson}} Arrays},
  author = {Wiesenfeld, Kurt and Colet, Pere and Strogatz, Steven H.},
  year = {1998},
  month = feb,
  journal = {Physical Review E},
  volume = {57},
  number = {2},
  pages = {1563--1569},
  publisher = {American Physical Society},
  doi = {10.1103/PhysRevE.57.1563},
  urldate = {2025-07-12}
}

@article{liMetastabilityMultipopulationKuramoto2025,
  title = {Metastability of Multi-Population {{Kuramoto}}--{{Sakaguchi}} Oscillators},
  author = {Li, Bojun and Uchida, Nariya},
  year = {2025},
  month = jan,
  journal = {Chaos: An Interdisciplinary Journal of Nonlinear Science},
  volume = {35},
  number = {1},
  pages = {013105},
  issn = {1054-1500},
  doi = {10.1063/5.0220321},
  urldate = {2025-07-29}
}

\clearpage
\onecolumngrid

\appendix
\section*{Supplemental Material}

The supplemental material is organized as follows: In Appendix~\ref{app:models}, we first explore the microscopic origins of the model studied in this work (Eq. \eqref{eq:OA}) via microscopic models such as disordered oscillators, non-reciprocal XY models, and flocking models. In \appref{app:eps}, we numerically demonstrate the occurrence of the critical exceptional point in different numbers of populations. In Appendix~\ref{app:z2restored}, we briefly provide a more general proof of the relationship between chiral motion and $\mathbb{Z}_2$ symmetry breaking. In Appendix~\ref{app:3pop}, we provide additional information on the SHO bifurcation and demonstrate the $N=3$ phase transitions on an alternative coupling network. In Appendix~\ref{app:npop}, we demonstrate the occurrence of crossovers within the chiral phase of in-phase to out-of-phase synchronized dynamics of sets of populations.

\section{Derivations of the \texorpdfstring{$N$}{N} Population Model}\label{app:models}
In this section, we discuss several models whose dynamics are described by Eq. \eqref{eq:OA}, beginning by reviewing the order parameter dynamics of Lorentzian-distributed disordered oscillators \cite{ottLowDimensional2008a,ottLongTimeEvolution2009}, following the derivation for multiple populations in \cite{liMetastabilityMultipopulationKuramoto2025}, then generalizing this derivation for arbitrary frequency distributions when the self coupling of each population is strong relative to the disorder. The same result is then found for XY spins and topological flocks.

\subsection{Kuramoto Oscillators}
\label{sec:ks_oscillators}
    The prototypical example admitting the equations of motion \eqref{eq:OA} is the multipopulation Kuramoto model. 
    Consider the system
    \begin{equation}
        \dot{\theta}_i^a = \omega^a_i + \sum_{b,j}\frac{J_{ab}}{N_b}\sin(\theta^b_j - \theta^a_i)\,,
        \label{eq:kuramotoosc}
    \end{equation}
    where $a$ and $b$ index populations and $i$ and $j$ index members of each population. 

The $n$-th Kuramoto-Daido order parameter for each population is defined as
\begin{equation}
    z_{a,n}(t) = \int_{-\infty}^\infty \int_0^{2\pi} P_a(\omega, \theta, t) e^{in\theta} \mathrm{d} \theta \mathrm{d} \omega, \label{eq:m-pop_op}
\end{equation}
where $P_a(\omega, \theta, t)$ is the probability density of an oscillator in population $a$ having frequency $\omega$ and phase $\theta$ at time $t$ in the thermodynamic limit. Conservation of probability implies each density obeys the continuity equation
\begin{equation}
    \frac{\partial P_a}{\partial t} + \frac{\partial}{\partial\theta}\mathcal{J}_a = 0, \label{eq:probflux}
\end{equation}
where $\mathcal{J}_a(\omega, \theta, t)$ is the probability current, which can be written using Eq. \eqref{eq:kuramotoosc} as
\begin{eqnarray}
    \mathcal{J}_a(\omega, \theta, t) &=& P_a(\omega, \theta, t)\left[ \omega + \frac{1}{2i}\sum_b J_{ab} \big(e^{-i\theta} z_b - e^{i\theta} \bar z_b \big)\right ],
\end{eqnarray}
where $z_a\equiv z_{a,1}$. Expanding $P_a(\omega, \theta, t)$ as a Fourier series in $\theta$ yields
\begin{equation}
    P_a(\omega, \theta, t) = \frac{g_a(\omega)}{2\pi} \left \{ 1 + \sum_{n=1}^\infty c_{a,n}(\omega, t) e^{in\theta} + \mathrm{c.c.} \right \}. \label{eq:m-population_KS_fourier}
\end{equation}
Using the Fourier expansion with Eq. \eqref{eq:m-pop_op}, $z_{a,n}$ can be written as
\begin{equation}
    z_{a,n}(t) = \int_{-\infty}^\infty g_a(\omega) \bar c_{a,n}(\omega, t) \mathrm{d} \omega.
    \label{eq:m-population_KS_op_OA}
\end{equation}
The Ott-Antonsen ansatz \cite{ottLowDimensional2008a}
supposes the relationship on the Fourier coefficients $c_{a,n}(\omega,t)=[c_{a,1}(\omega,t)]^n$. The ansatz results in the following dynamics on $\bar c_a\equiv \bar c_{a,1}$. 
\begin{equation}
    \dot{\bar{c}}_{a} = i\omega \bar c_{a} + \frac{1}{2}\sum_b
        J_{ab}(z_b - \bar c_{a}^2 \bar z_b).
    \label{eq:fourier_eq}
\end{equation}
A tractable choice for $g_a(\omega)$ is a Lorentzian distribution with center $\Omega_a$ and width $\Delta_a$ for each population,
\begin{equation}
    g_a(\omega) = \frac{1}{\pi}\frac{\Delta_a}{(\omega-\Omega_a)^2+\Delta_a^2}.
    \label{eq:m-population_KS_Lorentzian_def}
\end{equation}

The integral in \eqref{eq:m-population_KS_op_OA} can be evaluated after closing the contour in the lower half $\omega$-plane containing the pole at $\omega=\Omega_a-i\Delta_a$ yielding
\begin{equation}
    z_a(t) = \bar c_a(\Omega_a-i\Delta_a,t) .
    \label{eq:m-population_KS_op_integrated}
\end{equation}
This result allows us to convert Eq. ~\eqref{eq:fourier_eq} into the Ott-Antonsen equation for the mean-field dynamics,
\begin{equation}
    \dot{z}_a = -(\Delta_a-i\Omega_a) z_a+\frac{1}{2}\sum_b
    J_{ab}(z_b - z_a^2 \bar{z}_b).
    \label{eq:appOA}
\end{equation}

While the Ott-Antonsen equation holds for a Lorentzian frequency distribution, we will now show that it generally applies when populations of phase variables have sufficiently strong self-coupling in comparison to the frequency distribution width such that the populations are near synchrony. Let $\displaystyle g_{a}(\omega)
   =\frac{1}{\sigma_a}\,G\,\bigl(\tfrac{\omega-\Omega}{\sigma_a}\bigr)$ 
be any set of distributions which approach $\delta(\omega-\Omega)$ as $\sigma_a\to0$.
Assume $G(u)$ is normalized ($\int G(u)\,du = 1$), symmetric about $u=0$, has unit variance, and minimally has the first four moments finite for all populations. We will work in the rotating frame where $\Omega=0$ without loss of generality. 

For an arbitrary frequency distribution, the first two equations of motion in the Kuramoto-Daido hierarchy\cite{cestnikHierarchyExactLowDimensional2022} are

\begin{subequations}\label{eq:daidoarbfreq}
\begin{align}
\dot z_{a}
  &= 
   i \int_{-\infty}^{\infty} \omega \,g_a(\omega)\,
      \bar c_{a}(\omega)\,\mathrm{d}\omega
  + \frac{1}{2}\sum_b J_{ab}\!\left( z_{b} - \bar z_{b}\, z_{a,2} \right) \label{eq:daidoarbfreq:a}\\[4pt]
\dot z_{a,2}
  &= 
   2i\int_{-\infty}^{\infty} \,\omega\,g_a(\omega)
      \bar c_{a,2}(\omega)\,\mathrm{d}\omega
  + \sum_b J_{ab}\!\left( z_{b}\, z_{a} - \bar z_{b}\, z_{a,3} \right) \label{eq:daidoarbfreq:b}
\end{align}
\end{subequations}

The integrals in Eqs. ~\eqref{eq:daidoarbfreq} can be approximated in the limit of small width ($\sigma_a\to0$) as,

\begin{subequations}\label{eq:daidoint}
\begin{align}
\int \omega \,g_a(\omega)\,
      \bar c_{a}(\omega,t)\,d\omega
&= \sigma_a^2\,\bar c'_{a}(0,t)+O(\sigma_a^4),\label{eq:daidoint1}\\[4pt]
  \int \omega \,g_a(\omega)\,
      \bar c_{a,2}(\omega,t)\,d\omega
&= \sigma_a^2\,\bar c'_{a,2}(0,t)+O(\sigma_a^4)\approx 2\,\sigma_a^2\,\bar c_{a}(0,t)\,\bar c'_{a}(0,t)+O(\sigma_a^4),\label{eq:daidoint2}
\end{align}
\end{subequations}
where the final approximation in Eq. ~\eqref{eq:daidoint2} is a result of the Ott-Antonsen ansatz is violated to order $\sigma^2_a/J_{aa}$, and with $c'(0,t)$ being defined as $\partial_{\omega}c(\omega,t)$ evaluated at $\omega = \Omega=0$.
The equation of motion for $\bar c'_{a}(\omega,t)$ is
\begin{equation}
\dot{ \bar c}'_{a}(\omega)
= i\,\bar c_{a}(\omega)
+ i\,\omega\, \bar c_{a}'(\omega)
- \sum_b J_{ab}\, \bar z_b\,\bar c_{a}(\omega)\bar c'_{a}(\omega) +O(\sigma_a^2).
\end{equation}
Near the Ott-Antonsen manifold, $\kappa_{a,2}\equiv z_{a,2} - z_{a}^2$ is small,  allowing for the approximation $z_{a,3}\approx z_{a}^3 + 3z_{a}\kappa_{a,2}$. The dynamics of $\kappa_{a,2}$ is then given by
\begin{equation}
\dot\kappa_{a,2}
= 
 2i\sigma_a^2\bar c'_a(0)\big(2\bar c_a(0)-z_a \big)
- 2\,z_a\,\kappa_{a,2}\!\sum_b J_{ab}\bar z_b +O(\sigma_a^4).
\label{eq:kappa-ode}
\end{equation}
As the Ott-Antonsen manifold is approached in the limit of small width $g_a(\omega)$ such that each population is nearly synchronized, the variables $\kappa_{a,2}$ and $c'_{a}(0)$ implicit in Eq. \eqref{eq:daidoarbfreq:a} can be adiabatically eliminated. Performing the adiabatic elimination results in the following equation of motion for $z_a$,




\begin{equation}
\dot z_a
= 
 -\frac{\sigma_a^2}{2}\,\frac{z_a\,+\,O(\sigma_a^2)}{z_a\sum_bJ_{ab}\bar z_b}
 + \frac12\sum_b J_{ab}\Big(z_b - \bar z_b\,z_{a}^2\Big)
\label{eq:z-closedexact}
\end{equation}
Finally, because we assume the self-coupling of each populations is much greater than the frequency distribution width (i.e. $J_{aa}/\sigma_a^2\gg1$) and the interpopulation couplings, the form of the traditional Ott-Antonsen equation is recovered,
\begin{equation}
\dot z_a
= \left(i\Omega\,-\, \frac{\sigma_a^2}{2J_{aa}}\right) z_a
 \;+\; \frac12\sum_b J_{ab}\Big(z_b - \bar z_b\,z_{a}^2\Big)\;
 +\;O\left(\frac{\sigma_a^4}{J_{aa}}\right).
\label{eq:z-closed}
\end{equation}
where we used $|z_a|^2=1 -O(\sigma^2_a/J_{aa})$ and transformed back to the non-rotating frame.

\subsection{Non-reciprocal XY Models with all-to-all coupling}
The multipopulation all-to-all coupled XY model can be generalized to nonreciprocal systems in terms of the stochastic equations of motion,
\begin{equation}
    \dot{\theta}_i^a = \sum_{b,j}\frac{J_{ab}}{N_b}\sin( \theta^b_j-\theta^a_i) + \sqrt{2D_a}\eta^a_i,
    \label{eq:noisyXY}
\end{equation}
where $a,b,..$ index population and $i,j,..$ index the members of the population. Each $\eta^a_i$ are independent sources of Gaussian white noise satisfying $\langle\eta_i^a(t)\eta_j^b(t')\rangle = \delta_{a,b}\delta_{i,j}\delta(t-t')$ under the It\^{o} discretization. Similar to the result found in Ref. \cite{tyulkinaDynamicsNoisyOscillator2018}, the SDE for the order parameter, defined again as $z_{a,n} =  (1/N_a)\sum_i e^{in\theta^a_i}$ will satisfy 
\begin{equation}
    \frac{\mathrm{d}z_{a,n}}{\mathrm{d}t}=\sum_i \bigg((\partial_{\theta^a_i}z_{a,n})\dot{\theta}_i^a + D_a\partial^2_{\theta^a_i}z_{a,n}\bigg)
\end{equation}
by It\^{o}'s formula\cite{gardinerStochasticMethodsHandbook2009}. Using Eq. ~\eqref{eq:noisyXY}, we arrive at 
\begin{equation}
        \label{eq:noisyOA_hierarchy}
        \dot{z}_{a,n} = -n^2D_az_{a,n}+\frac{n}{2}\sum_{b} J_{ab}(z_{b}z_{a,n-1} - \bar z_{b}z_{a,n+1}) + in\sqrt{\frac{2D_a}{N_a}}\xi^a_n
    \end{equation}
    where 
    \begin{equation}
        \xi^a_n = \frac{1}{\sqrt{N_a}}\sum^{N_a}_ie^{in\theta_i}\eta^a_i.
    \end{equation}
    The noise source $\xi^a_n$ is complex noise satisfying $\langle \xi^a_n(t)\rangle = 0$ and $\langle \xi^a_n(t)\bar \xi^a_n(t')\rangle = \delta(t-t')$.
    In the limit $N_a\to \infty$, the coefficient of the stochastic term approaches zero and the system of equations becomes deterministic. We proceed analogously to the case of an arbitrary frequency disorder distribution.  In the small noise limit, the Ott-Antonsen ansatz is nearly satisfied, with  $\kappa_{a,2}\equiv z_{a,2} - z_{a}^2$ being small and $z_{a,3}\approx z_{a}^3 + 3z_{a}\kappa_{a,2}$.

    We have \begin{subequations}\label{eq:daido2XY}
\begin{align}
\dot{z}_{a} &= -D_az_{a}+\frac{1}{2}\sum_{b} J_{ab}\big(z_{b} - \bar z_{b}(z_{a}^2 + \kappa_{a,2})\big),
\\[4pt]
  \dot{\kappa}_{a,2}
  &= -2D_a(z_a^2+\,2\kappa_{a,2})
         \;-\;2z_a\,\kappa_{a,2}\sum_{b} J_{ab}\bar z_b.
\end{align}
\end{subequations}

In the weak-noise regime, \(\kappa_{a,2}\) relaxes quickly and can be adiabatically eliminated, yielding


\begin{equation}
\dot z_a
= -\,D_a z_a
+ \frac{1}{2}\sum_{b} J_{ab}\!\left[
  z_b
  - \bar z_b \left(
      z_a^{2}
      - \frac{ D_a z_a^{2} }{ 2D_a + z_a\sum_{c} J_{ac}\,\bar z_c  }
    \right)
\right].
\end{equation}

Assuming that $J_{aa}\gg D_a$ and $J_{aa}\gg |J_{ab}|$ for $b\neq a$, and using the fact that in the synchronized regime $|z_a|^2\gg2D_a/J_{aa}$, the form of the Ott-Antonsen equation is restored to leading order:
\begin{equation}
\dot z_a
=
-\,\frac{D_a}{2}\,z_a
+ \frac{1}{2}\sum_b J_{ab}\!\left(z_b - \bar z_b z_a^2\right).
\label{eq:Z_OA_limit}
\end{equation}


    \subsection{Flocking Models with Topological Interactions}
    Consider a flocking model with birds in the plane labeled by their position and angle pair $\{\vec{x}^i_a,\theta^i_a\}$, with flocks labeled by $a=A,B,C,\dots$ and members indexed by $i$. We take the microscopic dynamics to be given by 
    \begin{subequations}
    \begin{align}
        \dot{\theta}^i_a &= \sum_b\sum_{j\in\mathcal{N}^i_{ab}} \frac{J_{ab}}{N}\sin(\theta^j_b - \theta^i_a), \\
        \dot{\vec{x}}^i_a &= v\ \hat{r}(\theta^i_a),
        \label{eq:micro_flocking}
    \end{align}
    \end{subequations}
    where $\hat{r}(\theta^i_a)$ is the unit vector pointing in the direction $\theta^i_a$, and $\mathcal{N}^i_{ab}$ is the set of $M_a$ nearest birds of flock $b$ to bird $i$ in flock $a$, which we take to be the same for all interacting flocks $b$. For simplicity, we take all flocks to be of the same size $N$. This manner of couplings agents via topological distance rather than metric distance has been observed in flocks of starlings \cite{balleriniInteractionRulingAnimal2008}, and studied theoretically in a few contexts \cite{kumarEfficientFlockingMetric2021}. In the limit when the population sizes are large and when all birds see all birds, the dynamics of $z_a = R_ae^{i\phi_a} =(1/N)\sum_i e^{i\theta^a_i}$, becomes
    \begin{equation}
    \dot{z}_a = \frac{1}{2}\sum_b
    J_{ab}(z_b - z_a^2 \bar{z}_b),
    \label{eq:birdOAT0}
    \end{equation}
    the Ott-Antonsen equation with $\Delta_a = 0$. For sufficiently large attractive intrapopulation couplings in comparison to the cross-population coupling, Eq. \ref{eq:birdOAT0} yields a stable solution where 
    \begin{subequations}
    \begin{eqnarray}
        R_a(t) &=& 1,\\
        \dot \phi_a &=& \sum_b J_{ab}\sin(\phi_b - \phi_a),
        \label{eq:T0OA}
    \end{eqnarray}
    \end{subequations}
    where $R_a=1$ implies each population is perfectly synchronized. 
    Expanding around this limit, each flock member sees $M_a\coloneqq N-n_a$ nearest neighbors of each flock for an integer $n_a\ll N$.
    The order parameter dynamics becomes
    \begin{equation}
        \dot{z}_a=\frac{1}{2} \sum_b J_{a b}\left[z_b-z_a^2\bar{z}_b-\frac{1}{N^2}\sum_i\sum_{j \in \overline{\mathcal{N}}_{a b}^i} \bigg(e^{i \theta_j^b}-  e^{i(2\theta_i^a - \theta_j^b)}\bigg)\right]\,,
        \label{eq:flock_exact}
    \end{equation}
    where $\overline{\mathcal{N}}_{ab}^i$ is the complement of the set $\mathcal{N}^i_{ab}$. Near the synchronized limit, the excluded members of $\overline{\mathcal{N}}_{ab}^i$ can be well approximated by the mean of the population, resulting in
    \begin{equation}
        \dot{z}_a=\frac{1}{2} \sum_b J_{a b}\big(1-\frac{n_a}{N}\big)\big(z_b-z_a^2\bar{z}_b-A_{ab}\big),
        \label{eq:flock_Aab}
    \end{equation}
    where $A_{ab}=\langle e^{i(2\theta_i^a - \theta_j^b)} \rangle-\langle e^{i\theta_i^a}\rangle^2 \langle e^{ -i\theta_j^b} \rangle=\left(\langle e^{i(2\theta_i^a)} \rangle-\langle e^{i\theta_i^a}\rangle^2\right)\langle e^{ -i\theta_j^b} \rangle\equiv-\delta_az_a^2\bar z_b$, where $\delta_a= 1-\frac{\langle e^{i2\theta_a}\rangle}{z_a^2}$ is zero when the Ott-Antonsen ansatz is satisfied (when $M_a=N$), and $\langle\dots\rangle$ is the average over members. The mean-field dynamics in terms of $\delta_a$ becomes
    \begin{equation}
        \dot{z}_a=\frac{\delta_a}{2(1-\delta_a)} \sum_b \tilde J_{a b}z_b+\frac{1}{2} \sum_b \tilde J_{a b}\big(z_b-z_a^2\bar{z}_b\big)
        \label{eq:flock_meanfield}
    \end{equation}
    with $\tilde J_{ab}=(1-\delta_a)J_{ab}(1-\frac{n_a}{N_b})$.

    As an example, assume a von Mises phase distribution for each population, defined as
    
\begin{equation}
  P_a(\theta_i^a)
  = \frac{1}{2\pi\, I_0(\kappa_a)}\,
    \exp\!\big[\kappa_a \cos(\theta_i^a - \phi_a)\big],
  \label{eq:vm-a}
\end{equation}
    where $I_n(\kappa)$ is the modified Bessel function of the first kind of order $n$ \cite{NIST_DLMF}. In the limit of small width, this leads to
\begin{equation}
  \delta_a
  = 1 - \frac{I_2(\kappa_a)\, I_0(\kappa_a)}{I_1(\kappa_a)^2}
  \;\simeq\; \frac{1}{\kappa_a}
  \quad (\kappa_a \gg 1).
  \label{eq:deltaa-vm-leading}
\end{equation}

\section{Critical Exceptional Points}
\label{app:eps}

As mentioned in the main text, the non-reciprocity in the model creates a generic mechanism for exceptional point phase transitions. Here, we numerically verify the existence of the critical exceptional point at the transition from aligned to $N$-chiral phases for $N=3,4,5$ in Eq. ~\eqref{eq:OA}. The occurrence of an exceptional point can be determined for arbitrary periodic dynamics in a nonlinear system via a Floquet analysis of the limit cycle, as discussed in ~\cite{weisExceptionalPointsNonlinear2023a}. 

First consider a generic dynamical system,
\begin{eqnarray}
    \dot \phi_a=f_a({\phi_b}),
\end{eqnarray}
which exhibits a limit cycle $\phi_a(t)$ with period $T$. The time evolution operator $U(t-t_0)$ along a limit cycle
with period $T$ can be defined as the solution to the differential equation
\begin{align}
    \dot U(t-t_0)
  &= J(t-t_0)\,U(t-t_0),\\[1ex]
U(0)&=\mathbb{I},
\end{align}
where the components of the Jacobian $J(t-t_0)$ evaluated on a limit cycle $\phi_{LC}(t)$ are given by 
\begin{equation}
    J_{ab}(t-t_0) = \left. \frac{\partial f_a}{\partial \phi_b}\right|_{\phi=\phi_{LC}(t-t_0)}.
\end{equation} 
The operator $U(t-t_0)$ time evolves a vector in the tangent space of $\phi_{LC}(t_0)$ along the trajectory. 
Floquet vectors (a special case of the covariant Lyapunov vectors \cite{Ginelli2013})
for the limit cycle are given by the eigenvectors of $U(T)$, and the corresponding eigenvalues $\mu_i$ are related to the Lyapunov exponents via $\lambda_i = {\rm Re}[\ln(\mu_i)]/T$. The existence of a limit cycle ensures at least one zero Lyapunov exponent. 
When approaching a critical exceptional  point, a second Lyapunov exponent approaches zero and the two corresponding Floquet eigenvectors align.

We use this procedure to compute the angle between the first two Floquet eigenvectors $\theta_{12}$ in  Fig. \ref{fig:345EP}
as the parameter $J_-$ (defined in the figure caption) is tuned through the exceptional point phase transition. At the transition between the aligned and chiral phase for $N=3,4,5$ indicated by the dashed line, the angle between the two eigenvectors is zero.
\begin{figure*}[t]
    \includegraphics[width=\textwidth]{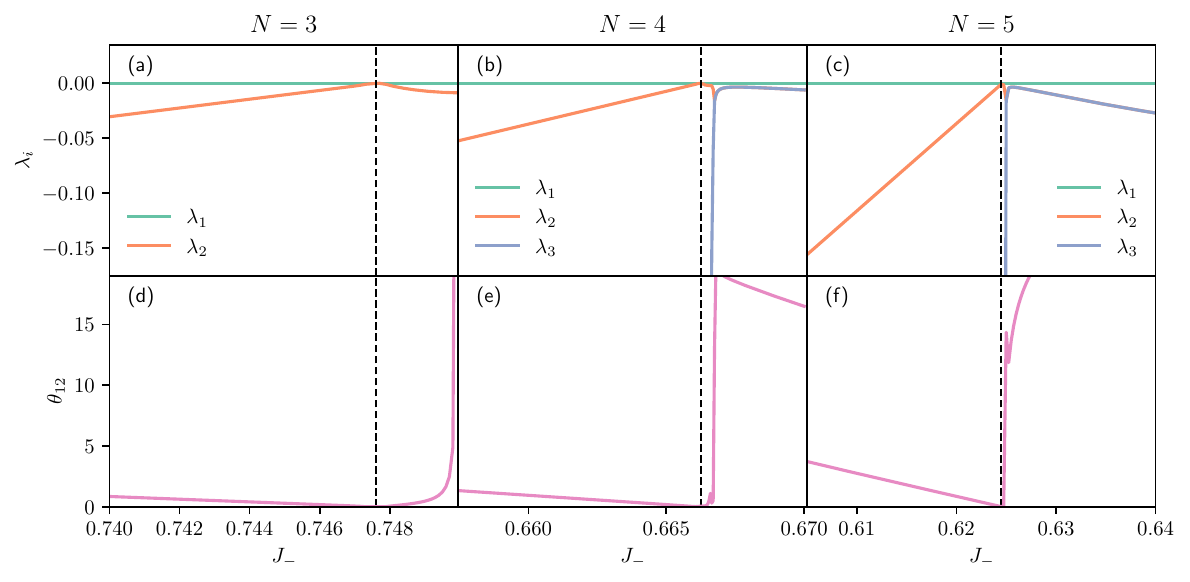}
    \caption{(a-c) Plots of the Lyapunov exponents $\lambda_i$ and (d-f) the angle between the Floquet vectors $\theta_{12}$ corresponding to the two lowest $\lambda_i$ for the aligned to chiral phase transition for $N=3,4,5$ with the phase transition demarcated by a dashed line occurring at $J_- = (0.7476, 0.66625, 0.6245)$ for each respective panel. The couplings in Eq. ~\eqref{eq:OA} are chosen to be $J_{ij} = 10\mathbb{I} + J_-\mathcal{S}_{ij} + (1-J_-)\mathcal{A}_{ij}$ with $\Delta=0.1$.}
    \label{fig:345EP}
\end{figure*}

\section{Symmetry Restored Phases}
\label{app:z2restored}
In this section, we prove the result from the main text 
that the total phase $\Phi$ does not exhibit net chiral motion for a generic $\mathbb{Z}_2$-symmetric attractor, i.e., 
\begin{eqnarray}
     \avgz{\dot\Phi}= 0, 
     \label{eq:finalresult}
\end{eqnarray} 
where $\avgz{~\cdot~}$ is the average of a quantity over the points on a trajectory $z_*(t)$ which satisfies the equations of motion Eq.~\eqref{eq:OA}. The $\mathbb{Z}_2$ symmetry enforces that the symmetry map will either take an attractor back into itself or to a symmetric partner. 
We refer to an attractor which is mapped into itself as $\mathbb{Z}_2$-symmetric attractor, meaning the  $\mathbb{Z}_2$ operation is a bijective map of the set of points into itself, which extends the definition to arbitrary chaotic or quasiperiodic solutions.
The dynamics of the total phase $\Phi$ is dictated by the equation 
\begin{equation}
     \dot{\Phi} = \sum_{ab} J_{ab}\bigg(\frac{R_b}{R_a}+R_aR_b\bigg)\sin(\phi_b - \phi_a)\equiv\sum_{ab}f_{ab}(\phi_{ab}).
     \label{eq:Phidynamics}
\end{equation}
The $\mathbb{Z}_2$ operation takes each phase $\phi_a\to-\phi_a$, and therefore $\Phi \to -\Phi$,  which leaves Eq. ~\eqref{eq:Phidynamics} invariant, allowing for the occurrence of such $\mathbb{Z}_2$-symmetric attractors. 

By applying the $\mathbb{Z}_2$ operation to a $\mathbb{Z}_2$ symmetric attractor $z_*(t)$, it follows that
\begin{eqnarray}
     \avgz{\dot\Phi}=\avgz{-\dot\Phi}=-\avgz{\dot\Phi}. 
\end{eqnarray} 
The first equality follows from the definition of the $\mathbb{Z}_2$-symmetric attractor implying that the average is over the same points before and after the reflection. 
Eq. ~\eqref{eq:finalresult} follows directly from the above equalities. 
This also serves as a definition of a $\mathbb{Z}_2$-symmetric or a $\mathbb{Z}_2$-broken but dynamically restored state. 


\section{Three Population Phase Diagram and Phase Transitions}
\label{app:3pop}
\begin{figure}[t]
    \centering
    \includegraphics[width=\textwidth]{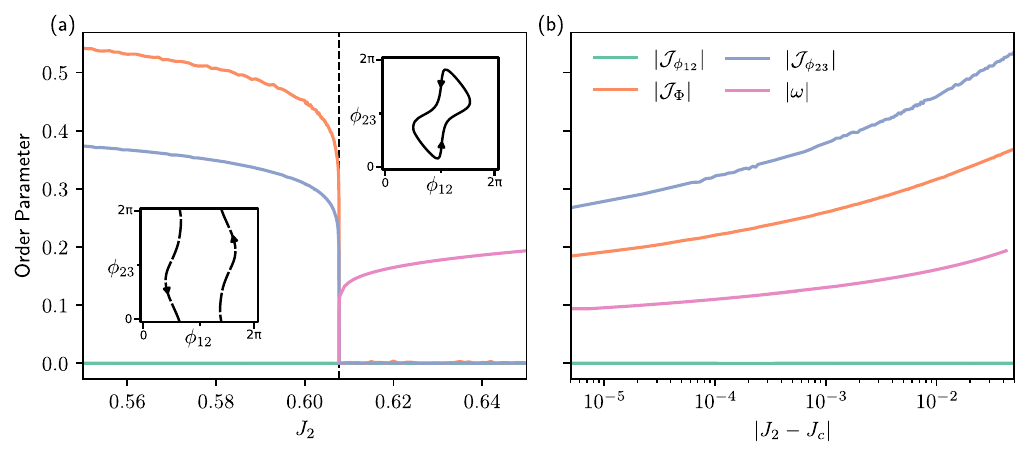}
    \caption{(a,b) The phase current order parameters as defined in the text and the limit cycle frequency $\omega$ (plotted on the periodic side of the phase transition) for identical parameters as the example studied in Fig. \ref{fig:z2hodynamics}. The phase transition occurs at $J_2=J_c\approx0.60774$.}
    \label{fig:Z2SHO_ops}
\end{figure}

Instead of computing the winding number defined in Eq. \eqref{eq:windingab}, which requires parameterizing the trajectory over one period, 
we find it practically more convenient to
compute the \emph{phase currents} to create the phase diagram in  Fig. \ref{fig:n3phasediagram}. We define the phase currents as 
\begin{subequations}
\begin{align}
\mathcal{J}_{\phi_{ab}} &= \lim_{T\rightarrow \infty}\frac{1}{T}\int_{t_0}^{t_0+T} \dot\phi_{ab}(t)\, dt,\\[1mm]
\mathcal{J}_{\Phi} &= \lim_{T\rightarrow \infty}\frac{1}{T}\int_{t_0}^{t_0+T}\dot\Phi\, dt,
\label{eq:phasecurrents}
\end{align}
\end{subequations}
with $\phi(t_0)$ being a point on the attractor and where $T$ is taken to be a finite but large number. The phase current is zero if the phase coordinate for the limit cycle does not wind around phase space in this direction, and otherwise has the same sign as the winding number (i.e., $\mathcal{J}=(-1,0,1)$ determines the chirality of winding around each coordinate). To further distinguish the non-winding trajectories, we compute the limit cycle center 
\begin{equation}
    \bar{\Delta\phi_{ab}}=\lim_{T\rightarrow \infty}\frac{1}{T}\int^{t_0+T}_{t_0}\phi_{ab}(t) dt,
\end{equation}
which yields $\bar{\Delta\phi_{ab}}= (0,0)$ or $(\pi,\pi)$ for Fig. \ref{fig:bifurcations}(d),(e) respectively. These two quantities are sufficient to distinguish the phase transitions found for $N=3$ populations.
Using the values of $\mathcal{J}_{\phi_{ab}}$ and $\bar{\Delta\phi_{ab}}$ on each periodic orbit at each point in parameter space produces Fig. \ref{fig:n3phasediagram}, where the configuration of trajectories corresponding to each color and number combination is shown in the surrounding panels.
\begin{figure}
    \centering
    \includegraphics[width=7in]{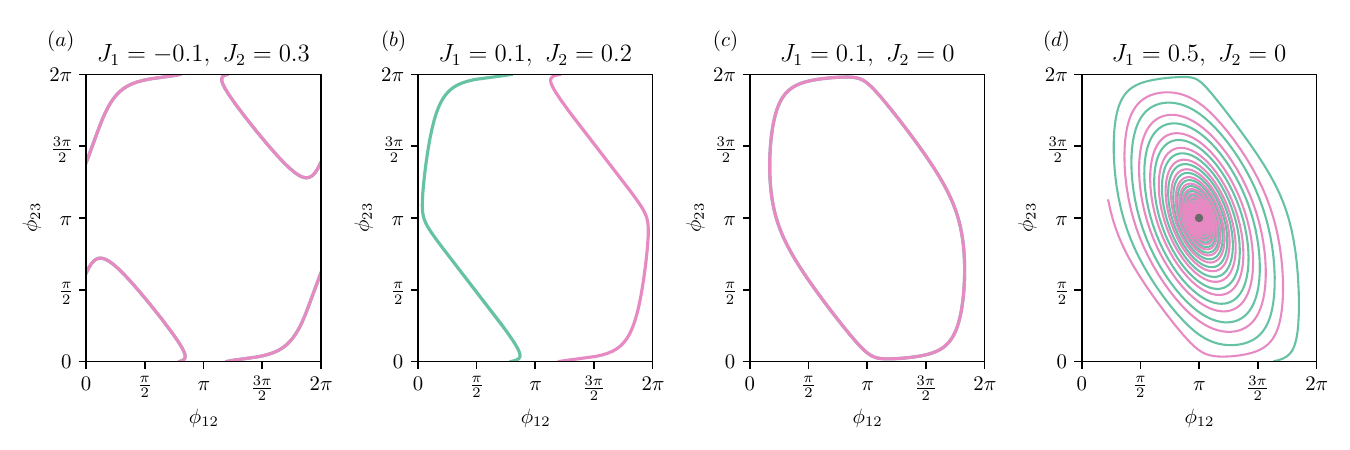}
    \caption{Identical phases as demonstrated for Fig. \ref{fig:n3phasediagram} for $J_{ij}$ defined in Eq.~\eqref{eq:transcoups} with $\Delta = 0.5$, which admits a coupling network with ``predator-prey" connectivity. In (d), the stable fixed point is shown as a gray dot.}
    \label{fig:trans_phases}
\end{figure}

The phase current continuously approaches zero at phase transitions, providing an order parameter for the phase transitions investigated here. Near a saddle homoclinic orbit bifurcation, the period is expected to diverge as 
\begin{equation}
    T\sim\frac{1}{\lambda_-}\ln|\mu - \mu_c|
\end{equation}
for a bifurcation occurring at $\mu = \mu_c$ and a saddle with unstable eigenvalue $\lambda_-$\cite{Kuznetsov2004}. In Fig.~\ref{fig:Z2SHO_ops}, we plot the limit cycle frequency on the symmetry-restored side of the phase transition, and $\mathcal{J}_{\phi_{ab}}$ and $\mathcal{J}_{\Phi}$ in the 2-chiral phase.

The existence of the various types of periodic orbits is not strongly dependent on the precise choices of parameters. For the more typical ``predator-prey" type couplings matrix below, 
\begin{equation}
    J_t = \begin{bmatrix} 
    10 & 1.3 & 0.7 +J_1 &\\
    -1.5 & 10 & 1.0+J_2 &\\
    -1.4 & -1.2 & 10
    \end{bmatrix}
    \label{eq:transcoups}
\end{equation}
we find a similar set of bifurcations as were shown in the main text. From the panels (a) to (b) and (b) to (c) in Fig. \ref{fig:trans_phases}, the $\mathbb{Z}_2$ SHO bifurcation occurs between the two non-winding cycles. From panels (c) to (d), a subcritical Hopf bifurcation also occurs, which was not found in the parameter region shown in Fig. \ref{fig:n3phasediagram}.

\section{In-phase and Out-of-Phase Synchrony within the \texorpdfstring{$N$}{N}-Chiral Phase}
\label{app:npop}

Within the chiral phase, multiple non-reciprocally coupled populations (Eq.  ~\eqref{eq:OA}) can admit dynamics where clusters of populations undergo transitions from in-phase to out-of-phase synchronization for $N\geq5$. We demonstrate this phenomenon in Fig. \ref{fig:nspecchiral}, where the aligned phase (red) indicates that all populations are perfectly in-phase, while in the remaining phases only  certain pairs of populations are in-phase. To produce the phase diagram, we take the couplings to be of the form 
\begin{equation}
J_{ab} =J_s\mathbb{I}_{ab} + \frac{J_{+}}{N}\mathcal{S}_{ab} + \frac{J_{-}}{N}\mathcal{A}_{ab}
\end{equation}
where $\mathcal{S}_{ab} = 1$ for $a\neq b$ and $\mathcal{S}_{ab} = 0$ for $a= b$, $\mathbb{I}_{ab}$ is the identity matrix, and 
\begin{equation}
\mathcal{A} = \begin{pmatrix}
0      & 1      & 1 & 1& \cdots\\
-1     & 0      & 1  & 1    & \cdots\\
-1     & -1     & 0    & 1  & \cdots\\
-1     & -1     & -1    & 0  & \cdots\\
\vdots & \vdots & \vdots & \vdots &\ddots \\ 
\end{pmatrix}.
\label{eq:transitive}
\end{equation}
In the orange portion of the chiral phase, three clusters of in-phase populations appear. We will refer to a cluster of one or more aligned populations as a synchrony group. The three synchrony groups in the orange phase chase each other, creating chiral dynamics. A transition in the number of synchrony groups occurs from the orange region to the yellow region, where an additional fourth group appears that is out-of-phase with the remaining three. Four synchrony group transitions occur while approaching the perfectly non-reciprocal limit for $N=7$ populations, where every pair of populations is out-of-phase. The synchrony group transitions shown in Fig. \ref{fig:nspecchiral}(b) appear sharp at small disorder (i.e. $\Delta\to0$), but as $\Delta$ increases, the transitions progressively soften, becoming apparent as smooth crossovers, as illustrated in Fig.~\ref{fig:nspecchiral}(c). The synchrony group transitions appear to become true phase transitions only in the perfect disorder-free limit $\Delta\to0$.

\begin{figure*}
    \includegraphics[width=\textwidth]{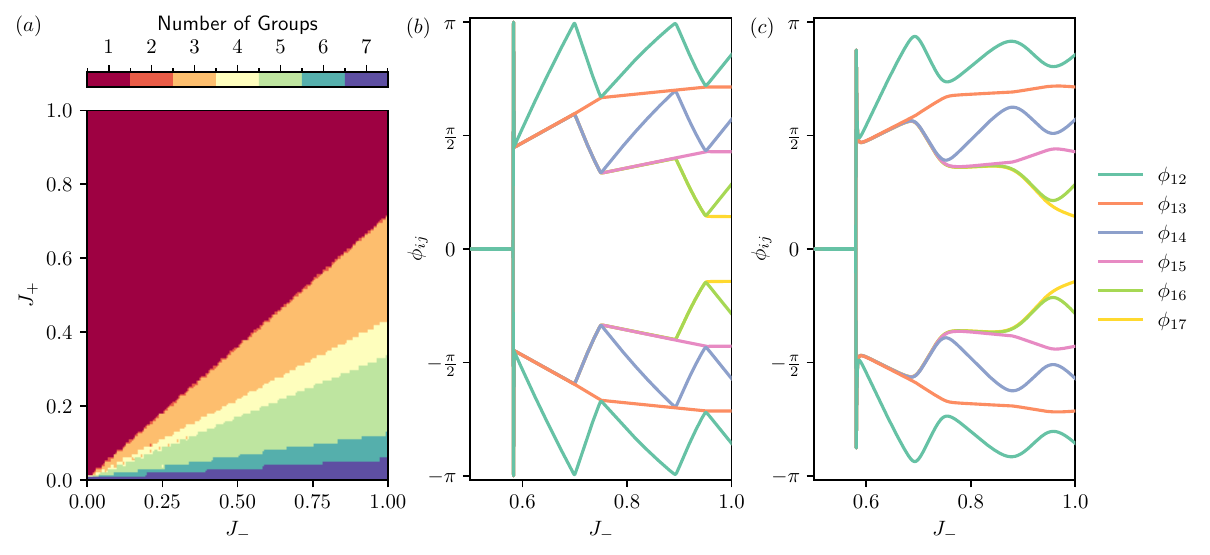}
    \caption{(a) Numerical Phase diagram for $N=7$ populations, where the color represents the number of synchrony groups (clusters of nearly aligned populations), for $\Delta = 0.25$, and $J_{ij} = 10\mathbb{I} + (J_{-}\mathcal{A} + J_{+}\mathcal{S)}/N$. The phase differences relative to the first population are plotted as a function of $J_-$ with $J_+=1-J_-$ (i.e. tuning along the diagonal of the phase diagram), with $\Delta=0.25$ in (b) and $\Delta=2.5$ in (c).}
    \label{fig:nspecchiral}
\end{figure*}

\end{document}